\documentclass{aa}  
\usepackage{graphicx}
\usepackage{txfonts}
\usepackage{hyperref}
\usepackage{lscape}
\usepackage{threeparttable}
\usepackage{xcolor}
\usepackage{multirow}
\usepackage{siunitx}
\usepackage[normalem]{ulem}
\usepackage{orcidlink}
\usepackage{booktabs}
\usepackage{soul}

\begin{document} 

\title{JWST imaging of omega\,Centauri -- I.}
\subtitle{Luminosity and mass functions of its main sequence populations}

\author{M.\,Scalco\inst{1}\orcidlink{0000-0001-8834-3734}\fnmsep\thanks{\email{mscalco@iu.edu}}
\and 
R.\,Gerasimov\inst{2}\orcidlink{0000-0003-0398-639X}
\and
L.\,R.\,Bedin\inst{3}\orcidlink{0000-0003-4080-6466}
\and
E.\,Vesperini\inst{1}
\and
D.\,Nardiello\inst{4,3}\orcidlink{0000-0003-1149-3659}
\and
M.\,Libralato\inst{3}\orcidlink{0000-0001-9673-7397}
\and
A.\,Burgasser\inst{5}\orcidlink{0000-0002-6523-9536}
\and
M.\,Griggio\inst{6}\orcidlink{0000-0002-5060-1379}
\and
A.\,Bellini\inst{6}
\and
J.\,Anderson\inst{6}
\and
M.\,Salaris\inst{7,8}\orcidlink{0000-0002-2744-1928}
\and
D.\,Apai\inst{9,10}\orcidlink{0000-0003-3714-5855}
\and
M.\,H\"{a}berle\inst{11}\orcidlink{0000-0002-5844-4443}
}

\institute{
Department of Astronomy, Indiana University, Swain West, 727 E. 3rd Street, Bloomington, IN 47405, USA
\and
Department of Physics and Astronomy, University of Notre Dame, Nieuwland Science Hall, Notre Dame, Indiana 46556, USA
\and
Istituto Nazionale di Astrofisica, Osservatorio Astronomico di Padova, Vicolo dell’Osservatorio 5, Padova I-35122, Italy
\and
Dipartimento di Fisica e Astronomia "Galileo Galilei", Universit{\`a} di Padova, Vicolo dell'Osservatorio 3, Padova I-35122, Italy
\and
Department of Astronomy \& Astrophysics, University of California, San Diego, La Jolla, California 92093, USA
\and
Space Telescope Science Institute, 3700 San Martin Drive, Baltimore, MD 21218, USA
\and
Astrophysics Research Institute, Liverpool John Moores University, 146 Brownlow Hill, Liverpool L3 5RF, UK
\and
Istituto Nazionale di Astrofisica, Osservatorio Astronomico d'Abruzzo, Via Mentore Maggini, Teramo I-64100, Italy
\and
Department of Astronomy and Steward Observatory, The University of Arizona, 933 N. Cherry Avenue, Tucson, AZ 85721, USA
\and
Lunar and Planetary Laboratory, The University of Arizona, 1629 E. University Blvd., Tucson, AZ 85721, USA
\and 
Max Planck Institute for Astronomy, K\"onigstuhl 17, D-69117 Heidelberg, Germany
}

\date{XXX,YYY,ZZZ}
 
\abstract
{This paper presents the first study of the most massive globular cluster (GC) in the Milky Way, omega\,Centauri (or $\omega$\,Cen, also known as NGC\,5139), employing recently acquired JWST deep images. By combining these data with archival \textit{Hubble} Space Telescope (HST) images, we derived proper motions (PMs) for a significant portion of the JWST field. Our analysis of the colour-magnitude diagram (CMD) reveals two prominent sequences extending from a magnitude $m_{\rm F322W2} \sim$17.5 to the bottom of the main sequence (MS). These sequences correspond to the two main stellar populations of omega\,Centauri: the bMS (He-rich) and rMS (He-normal) populations. The two sequences intersect at the MS knee ($m_{\rm F322W2} \sim 19.5$) and change positions for lower magnitudes, with the bMS luminosity function (LF) ending at least $\sim$0.5 magnitudes brighter than the rMS LF. Our comparison with theoretical isochrones shows that the colour spread in the CMD is primarily driven by variations in helium abundance above the MS knee, while below the MS knee, the broader colour distribution is mainly influenced by variations in oxygen and carbon abundances, in combination with metallicity differences. We find that a single-population broken power-law mass function (MF) provides the best fit to the data. The MF exhibits a break around $0.2\ \mathrm{M}_\odot$, with a steep slope above the break and a flatter slope below it. Finally, we identified a third group of stars (named gMS) along the main sequence located between the two primary ones and conducted a detailed analysis of the LFs and MFs for these three stellar populations. The LFs of these sequences show similar trends, with the rMS being the most populated and the bMS the least. The MFs display distinct power-law slopes: the rMS is well fitted by a single power-law while the gMS and the bMS are characterised by MFs steeper than that of the rMS for masses larger than 0.2\,M$_{\odot}$ and flatter MFs for smaller masses. The flattening around $\sim$0.2\,M$_{\odot}$ for the gMS and the bMS might be a real feature of the MFs of these populations or due to uncertainties in the adopted mass-luminosity relationship (MLR). The variation in the slope of the MFs of the gMS and bMS contributes to the steepening (flattening) of the combined MF for masses higher (lower) than 0.2\,M$_{\odot}$. 
}

\keywords{globular clusters: individual: NGC\,5139, Proper motions, Stars: luminosity function, mass function}

\titlerunning{JWST view of omega\,Centauri}
\authorrunning{M.\,Scalco et al.}
\maketitle

\section{Introduction}\label{Section1}
Globular clusters (GCs), being the oldest and largest coeval stellar populations, serve as crucial markers for the chemical evolution of the Galaxy and provide valuable insights into the nature of ancient metal-poor stars. The long-standing belief that all stars within a cluster have the same chemical composition was challenged by spectroscopic studies, which revealed the presence of chemical inhomogeneities in some clusters \citep[see e.g.,][and references therein]{2012A&ARv..20...50G}. Two decades of space-based photometric studies have further revolutionised our understanding of GCs, uncovering intricate colour-magnitude diagrams (CMDs) that reveal the presence of multiple stellar populations (mPOPs) with varying chemical compositions. Despite these discoveries, the process driving this chemical diversity remains unclear, partly because most research on GCs has focused on the brighter stars, such as those on the upper main sequence (MS) or in the giant phase \citep[see][for a recent review]{mPOPs_review,2019A&ARv..27....8G}.

In this paper, we present JWST images of NGC\,5139 (hereafter referred to as $\omega$\,Cen), the relatively close \citep[we assumed the distance of $5.24\pm0.11$\,kpc from][throughout the paper]{2021ApJ...908L...5S} and most massive \citep[$\sim 4 \times 10^6$ M$_{\odot}$][]{2003MNRAS.339..486G,2013MNRAS.429.1887D} GC in the Milky Way. These observations aim to investigate the faintest objects within the cluster using near-infrared (NIR) data and explore the mPOPs in the faintest regions of its CMD. 

$\omega$\,Cen is well known for its complex mPOP system, making it one of the most enigmatic stellar systems in the Galaxy. It hosts at least two main stellar components: the blue main sequence (bMS) and the red main sequence (rMS) \citep[e.g.,][]{2004ApJ...605L.125B}, with the bMS more centrally concentrated than the rMS \citep{2009A&A...507.1393B,2007ApJ...654..915S}. Except for the cluster's central region, the rMS is the dominant population. This initially appeared inconsistent with spectroscopic studies, which indicated that lower-metallicity stars are the most abundant in the cluster \citep[e.g.,][]{1995ApJ...447..680N}. To resolve this discrepancy, \citet{2004ApJ...605L.125B} proposed that the bMS might be significantly enhanced in helium than the rMS. This hypothesis was supported by \citet{2004ApJ...612L..25N}, who estimated a helium mass fraction difference of $\Delta Y \sim 0.12$ between the two sequences. Further confirmation came from \citet{2005ApJ...621..777P}, who demonstrated that bMS stars are indeed more metal-rich than their rMS counterparts by $\sim$ 0.3\,dex, while \citet{2012AJ....144....5K} derived a helium mass fraction of $Y = 0.39 \pm 0.02$ for the bMS. However, a more recent spectroscopic analysis by \citet{2021A&A...653L...8L} based on MUSE data suggests a smaller helium difference between the bMS and rMS ($Y \lesssim 0.1$).

Another remarkable feature of $\omega$\,Cen is the unusually broad metallicity spread observed within each sequence, significantly larger than what is typically found in other globular clusters. The metallicity ranges from [Fe/H] $\approx$ $-$2.2 to $-$0.5, with [X/Fe] dispersions reaching 0.5 to more than 1.0 dex for several elements \citep[e.g.][]{1995ApJ...447..680N,1996AJ....111.1913S,2000AJ....119.1239S,2008ApJ...681.1505J,2017ApJ...844..164B,2009ApJ...698.2048J}. This suggests that $\omega$\,Cen might be the nucleus of a dwarf galaxy absorbed by the Milky Way, or the outcome of the merger of two or more clusters \citep{1997ApJ...487L.187N,1998ApJ...506L.113J,2003MNRAS.346L..11B,2000ApJ...534L..83P,2006ApJ...637L.109B,2019NatAs...3..667I,2006A&A...445..513V}. Recent studies have further revealed that both the bMS and rMS are subdivided into multiple subpopulations, with up to 15 distinct stellar populations identified within the cluster so far \citep{2017ApJ...844..164B,2024A&A...688A.180S}.

The high-resolution deep NIR images provided by JWST have proven to be very effective in the exploration of mPOPs in GCs (47\,Tuc \citealt{2023MNRAS.522.2429M,2024ApJ...969L...8M}, M92 \citealt{2022MNRAS.517..484N}, NGC\,6440 \citealt{2023A&A...679L..13C} and NGC\,6397 \citealt{2024A&A...689A..59S}). Here, we present the first\footnote{After our submission of this paper (on May 8, 2025), a manuscript on the dynamical properties of $\omega$\,Cen based in part on JWST observations was posted on arXiv (Ziliotto et al., June 26, 2025, \href{https://arxiv.org/abs/2506.21187v1}{arXiv:2506.21187v1}, submitted to A\&A).} analysis of the mPOPs in $\omega$\,Cen using JWST photometry, with a focus on their luminosity and mass function along the MS. 

The paper is organised as follows: Section\,\ref{Section2} describes the data and reduction methods, while Section\,\ref{Section3} focuses on the evaluation of proper motions. Section\,\ref{Section4} presents the CMD of $\omega$\,Cen based on JWST photometry, followed by Section\,\ref{Section5}, which details the artificial star (AS) tests. Section\,\ref{Section6} analyses the luminosity and mass functions of $\omega$\,Cen and its stellar populations using JWST data. Finally, a brief summary is provided in Section\,\ref{Section7}.

\section{Observations and data reduction}\label{Section2}
The data analysed in this study were obtained from two programmes:
\begin{itemize}
    \item Our proprietary images from JWST GO-5110 programme \citep{2024jwst.prop.5110B}. The JWST data were obtained with the Near Infrared Camera \citep[NIRCam, ][]{2023PASP..135b8001R} simultaneously with the Short Wavelength (SW) and Long Wavelength (LW) channels on August 7-8, 2024 (epoch $\sim$2024.6). Ultra-wide filters were used for both NIRCam channels: F150W2 for the SW channel and F322W2 for the LW channel. The 6-point \texttt{FULLBOX} primary pattern was employed with a \texttt{3-POINT-LARGE-WITH-NIRISS} sub-pixel dither pattern. At each of the resulting 18 pointings, a single image was captured in both channels using the \texttt{MEDIUM8} readout pattern (9 groups), resulting in an effective exposure time of 944.836\,s per image.
    \item Archival material from our HST GO-14118+14662 \citep{2016hst..prop14118B,2016hst..prop14662B} multi-epoch programme. This dataset includes images taken with the Wide Field Channel (WFC) of the Advanced Camera for Surveys (ACS) using the F606W and F814W filters, spanning a period from $\sim2015.6$ to $\sim2018.5$. These data were reduced and presented in a recent paper by \citet{2024A&A...691A..96S}, where a comprehensive description of the dataset can be found.
\end{itemize}

\begin{centering} 
\begin{figure}
 \includegraphics[width=\columnwidth]{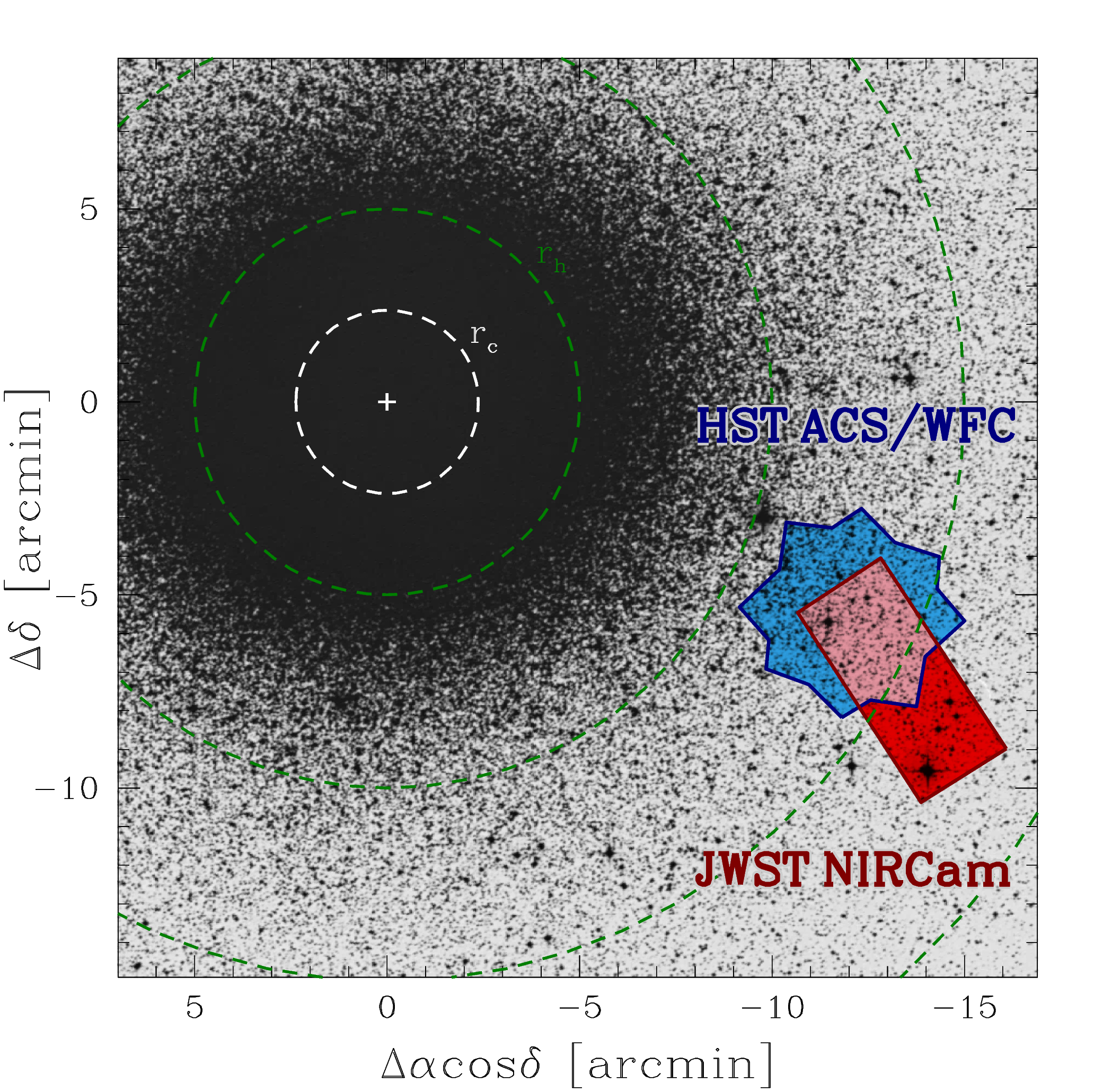}
 \caption{The NIRCam JWST field from the GO-5110 programme (red) and the ACS/WFC HST field from the GO-14118+14662 programme (blue) overlaid on a DSS image of $\omega$\,Cen. The overlap region between the two datasets is highlighted in magenta. Units are in arcminutes from the cluster's centre. The white and green dashed circles represent the core radius ($r_{\rm c}=2^\prime_{\cdot}37$) and the half-light radius ($r_{\rm h}=5^\prime_{\cdot}00$), respectively, with additional circles marking $2\,r_{\rm h}$, $3\,r_{\rm h}$, and $4\,r_{\rm h}$. The values of $r_{\rm c}$ and $r_{\rm h}$ are from \citet{1995AJ....109..218T,2005ApJS..161..304M} as reported in the \citet{1996AJ....112.1487H,2010arXiv1012.3224H} catalogue.} 
 \label{FOV} 
\end{figure} 
\end{centering}

Figure\,\ref{FOV} shows the positions of our datasets within the field of view (FOV), overlaid on a Digital Sky Survey\footnote{\href{https://archive.eso.org/dss/dss}{https://archive.eso.org/dss/dss}} (DSS) image of $\omega$\,Cen. The overlapping region between the two datasets, highlighted in magenta, covers a radial range from $\sim2.4$\,r$_h$ to $\sim3.2$\,r$_h$.

JWST images were processed using the software tools and methods detailed in Papers I, II, and III of the "Photometry and Astrometry with JWST" series \citep{2022MNRAS.517..484N,2023AN....34430006G,2023MNRAS.525.2585N}, which have been successfully applied in recent studies of 47\,Tuc by \citet{2023MNRAS.521L..39N,2025A&A...694A..68S}. We first processed the level-1b uncalibrated (\texttt{\_uncal}) images using a development version of the JWST pipeline\footnote{\href{https://github.com/spacetelescope/jwst}{https://github.com/spacetelescope/jwst}} \citep{2023BushouseJWSTpipeline}, running it through stages 1 and 2 to produce the level-2b (\texttt{\_cal}) images. In the stage 1 pipeline, we utilised the default parameters with one exception: for the ramp fitting process, we employed the \texttt{frame\,zero} (the initial frame of each integration) to measure pixels that were saturated in the first group and extended up to the ramp. This approach effectively increased the dynamic range of our data by almost two magnitudes, extending into the nominally saturated intensity regime. The stage 2 pipeline was executed with all default parameters.

The reduction of the \texttt{\_cal} images involved two main steps: first-pass and second-pass photometry. In the first-pass, starting from the empirical library Point Spread Function (PSF) derived in \citet{2024AN....34540039B}, we extracted PSFs, positions, and magnitudes of stars from each individual image. A geometric distortion correction was applied to the star positions using the solution by \citet{2023AN....34430006G}. The coordinates from each image were then transformed to a common reference frame, with bright cluster members from {\it Gaia} Data Release 3 \citep[DR3; ][]{2016A&A...595A...1G,2023A&A...674A...1G} serving as the reference, after transforming their positions to the epoch of the data collection.

For the second-pass photometry, we utilised a modified version of the \texttt{KS2} code as described in \citet{2017ApJ...842....6B,2018ApJ...853...86B,2018MNRAS.481.3382N,2018ApJ...854...45L,2022LibralatoPMcat,2021MNRAS.505.3549S} and references therein. In this step, positions and fluxes were extracted using the PSFs and transformations obtained during the first-pass. \texttt{KS2} processes all images simultaneously, making it well-suited for detecting faint sources that may not be visible in individual frames. Along with fluxes and positions, \texttt{KS2} generates several quality parameters, such as the PSF quality-of-fit (QFIT) parameter, the RADXS parameter \citep[a metric for comparing the shape of a source to the PSF; see][]{2008ApJ...678.1279B,2009ApJ...697..965B}, and the local sky noise (rmsSKY). Detailed descriptions of these diagnostics can be found in \citet{2018ApJ...853...86B,2021MNRAS.505.3549S,2018MNRAS.481.3382N}.

We applied a similar reduction process to the HST data, following the same first- and second-pass photometry approach. For a detailed explanation of the HST data reduction, we refer to \citet{2024A&A...691A..96S}.

We calibrated our HST and JWST photometry to the VEGA-magnitude system following the procedures illustrated in \citet[][see also \citealt{2005MNRAS.357.1038B}]{2023MNRAS.525.2585N}, while we anchored the astrometry to the International Celestial Reference System (ICRS) frame using \textit{Gaia} DR3 data for sources in the observed fields.

\begin{centering} 
\begin{figure}
 \includegraphics[width=\columnwidth]{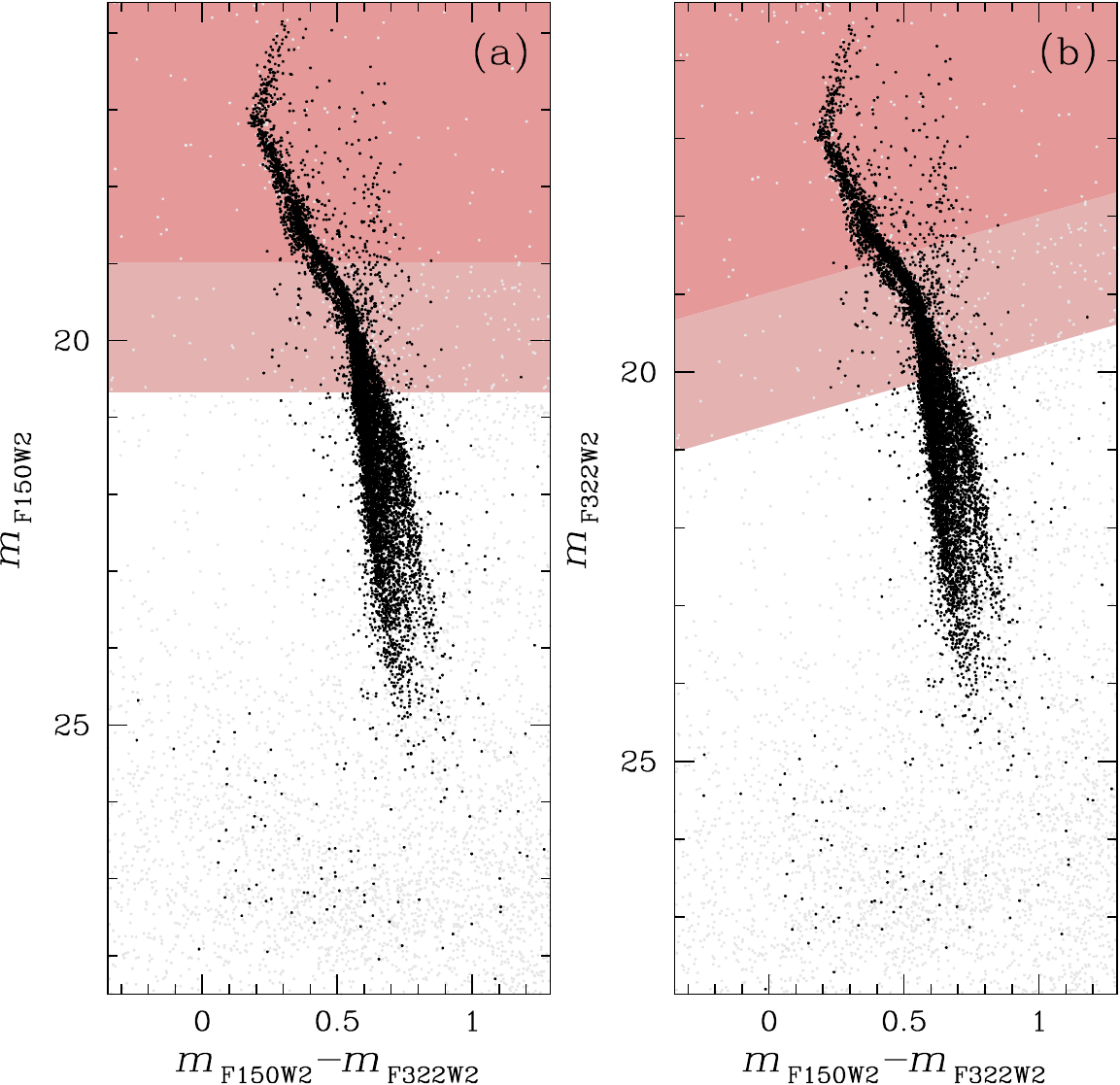}
 \caption{Colour-magnitude diagrams of $\omega$\,Cen using JWST filters. (a) $m_{\rm F150W2}$ versus $m_{\rm F150W2}-m_{\rm F322W2}$ CMD. (b) $m_{\rm F322W2}$ versus $m_{\rm F150W2}-m_{\rm F322W2}$ CMD. In both panels, black dots represent stars passing the quality selections while grey dots represent stars not passing the selections. The light-red and dark-red shaded regions highlight areas of the CMDs affected by saturation. The light-red region corresponds to saturated photometry, while the dark-red region represents severe saturation, where the photometry is saturated even in the \texttt{frame\,zero}.} 
 \label{CMDs} 
\end{figure} 
\end{centering}

We selected a sample of well-measured stars using the quality parameters provided by \texttt{KS2} (QFIT, RADXS, and rmsSKY). n particular, we retained sources with \texttt{QFIT} > 0 and absolute \texttt{RADXS} values smaller than 0.05 in both filters. As for \texttt{rmsSKY}, we manually defined a fiducial threshold as a function of magnitude, following the trend of \texttt{rmsSKY} with magnitude, and excluded all sources lying above this threshold. Finally, we corrected our photometry for the effects of differential reddening and spatial zero-point variations following the procedure introduced in \citet[][see also \citealt{2012A&A...540A..16M,2017ApJ...842....7B}]{2007AJ....133.1658S}. Filter-specific extinction coefficients were computed assuming $A_{\rm V} = 3.1 \times E(B-V)$, with a mean reddening $E(B-V) = 0.12$ \citep{1996AJ....112.1487H,2005ApJ...634L..69C}. We adopted the extinction ratios for JWST/NIRCam filters from \citet{2019ApJ...877..116W} ($A_{\rm F150W2} = 0.15 \times A_V$, $A_{\rm F322W2} = 0.04 \times A_V$). The measured differential reddening values $\delta E(B-V)$ range from approximately $-$0.05 to $+$0.04 across the field. Figure\,\ref{CMDs} presents the $m_{\rm F150W2}$ versus $m_{\rm F150W2}-m_{\rm F322W2}$ colour-magnitude diagram (CMD, panel a) and the $m_{\rm F322W2}$ versus $m_{\rm F150W2}-m_{\rm F322W2}$ CMD (panel b) for all stars in the JWST field. In both panels, black dots represent stars passing the photometric quality selections while grey dots represent stars not passing the selections. The black dots distinctly outline the cluster's main sequence (MS) and low-MS, as well as its white dwarf (WD) cooling sequence (CS), located in the bottom-left of the CMDs. In both panels, the light-red area indicates the onset of saturation, where photometry is obtained from the \texttt{frame\,zero}. The dark-red region marks severe saturation, where even the \texttt{frame\,zero} is affected and the flux must be recovered from the unsaturated wings of the PSF. In this region, the photometry becomes unreliable, with the sequence displaying non-physical features. Both the photometry and the astrometry of stars in these regions must therefore be treated with particular caution. Stars within these saturated regions are not ideal for scientific analysis that require high photometric accuracy -- such as the determination of the luminosity functions (LFs) and mass functions (MFs) of the individual stellar populations (see Section\,\ref{Section6.2}), which represents the core of this study --  and will not be utilised for these purposes. However, for completeness and to provide a comprehensive view of the CMD morphology of $\omega$\,Cen, we still include and make use of these stars in other parts of the paper. The reader should nevertheless keep in mind that any results involving stars in the saturated regime must be interpreted with appropriate caution.

\begin{centering} 
\begin{figure}
 \includegraphics[width=\columnwidth]{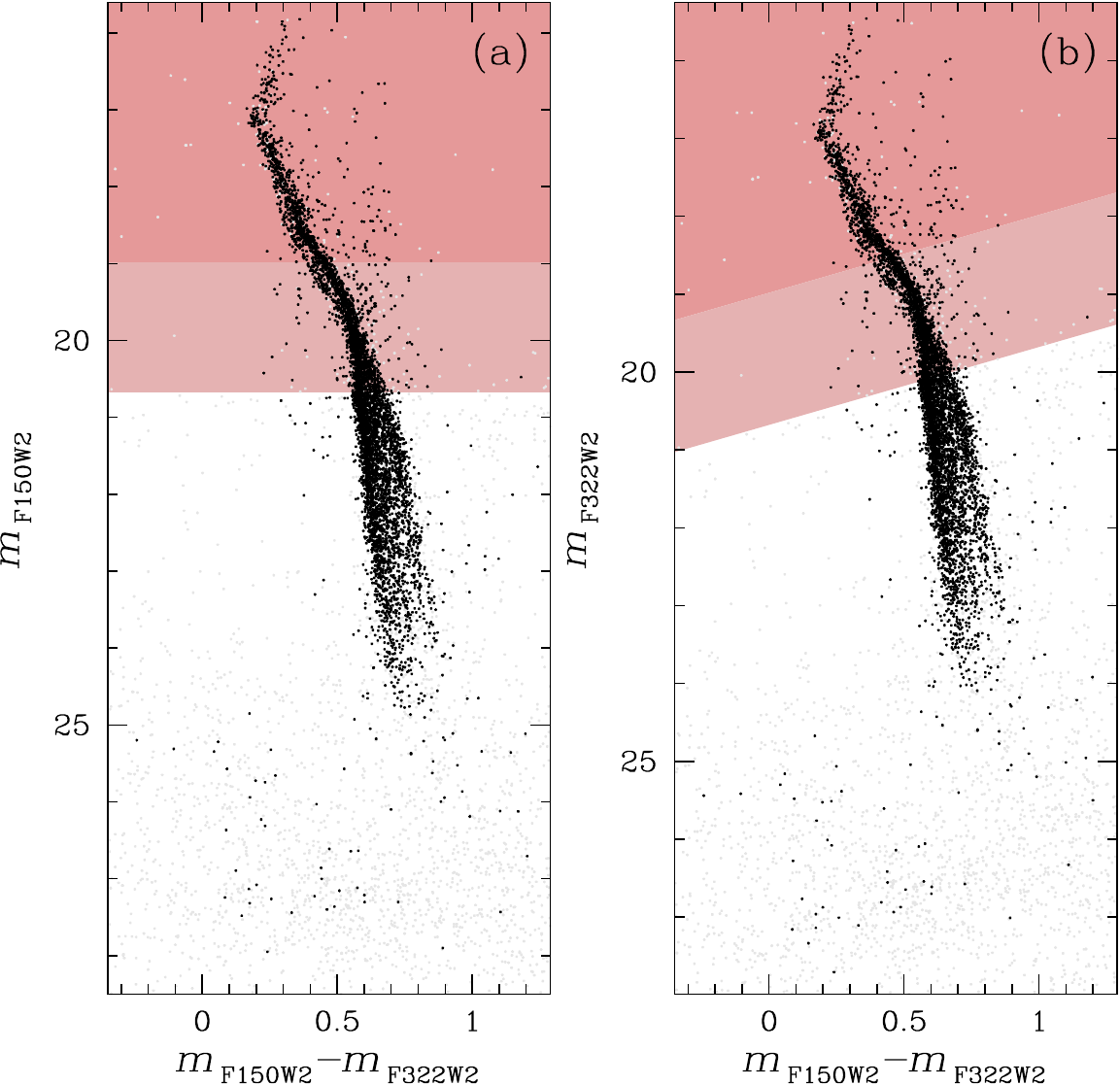}
 \caption{Same as Fig.\,\ref{CMDs}, but only for the sources located in the overlap region between the JWST and HST FOV.} 
 \label{CMDs2} 
\end{figure} 
\end{centering}

Figure\,\ref{CMDs2} shows the same CMDs as in Fig.\,\ref{CMDs}, but limited to stars located in the overlap region between the JWST and HST FOV. While the number of sources is noticeably reduced (by $\sim30$\%), the two sequences along the MS remain visible. However, the presence of some contaminants highlights the need for a proper motion-based cluster membership selection, which will be addressed in the following section.

\section{Proper motions}\label{Section3}
\begin{centering} 
\begin{figure*}
\centering
 \includegraphics[width=0.9\textwidth]{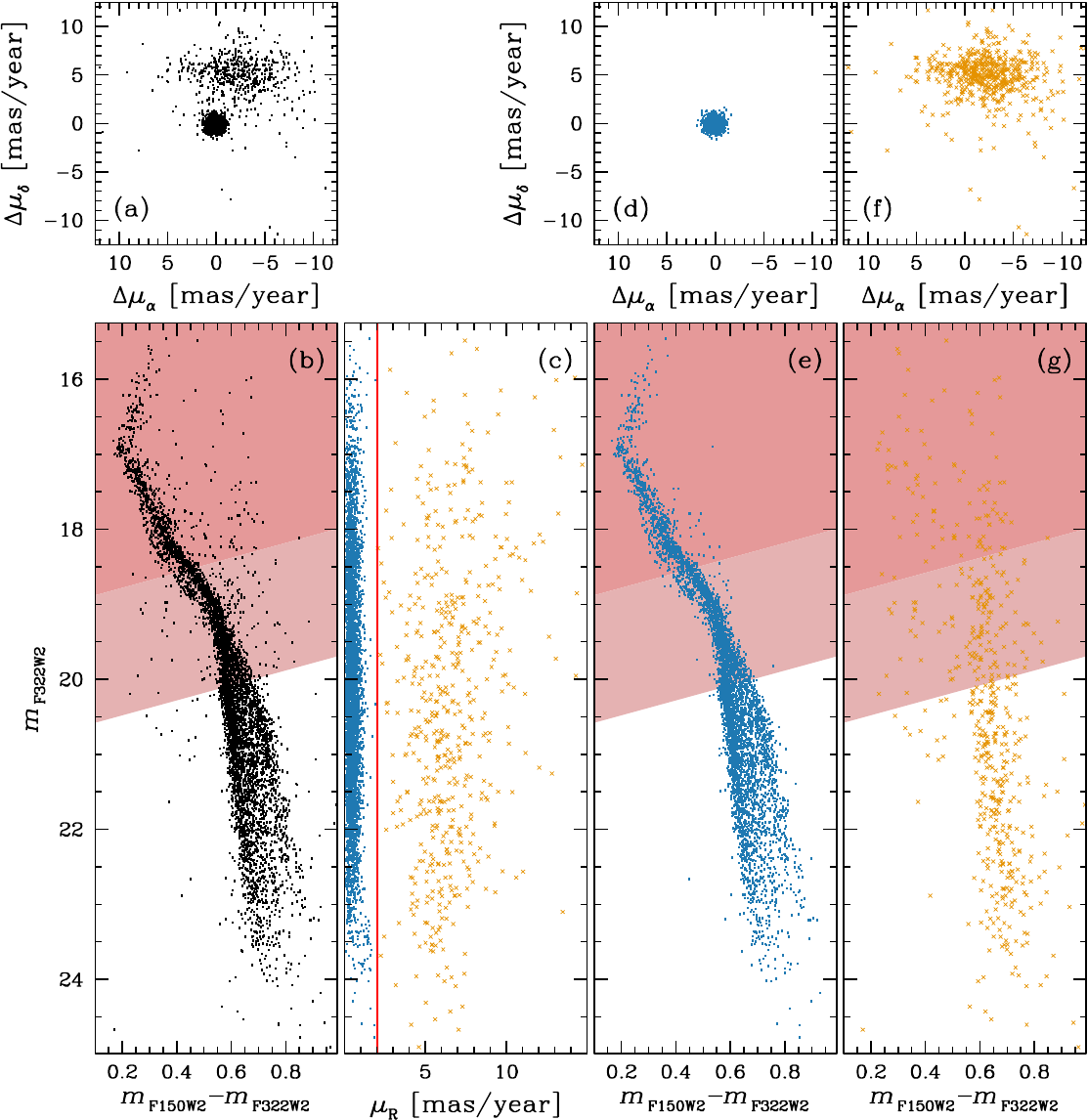}
 \caption{Cluster membership selection based on PMs, for stars that passed the photometric quality selections and have measurable PMs. (a)-(b) VPD and $m_{\rm F322W2}$ versus $m_{\rm F150W2}-m_{\rm F322W2}$ CMD, respectively. (c) $m_{\rm F322W2}$ magnitude versus the one-dimensional PM ($\mu_{\rm R}$). The red line separates cluster members from field stars. (d)-(e) VPD and $m_{\rm F322W2}$ versus $m_{\rm F150W2}-m_{\rm F322W2}$ CMD for stars that passed the PM selection. (f)-(g) same diagrams for stars that did not pass the PM selection. In panels (c) through (g), sources that passed the PM selection are represented by blue dots, while those that did not pass are depicted as orange crosses. In panels\,(b), (e) and (g), the light-red and dark-red shaded regions highlight areas of the CMDs affected by mild and severe saturation, respectively.}
 \label{PMs} 
\end{figure*} 
\end{centering}

Proper motions (PMs) were computed as displacements between the HST and JWST observations divided by the temporal baseline ($\sim$7.5\,years). For the HST epoch, we adopted the average position across the available exposures, given the much shorter time span relative to the HST–JWST baseline. The resulting PMs are shown in Fig.\,\ref{PMs}, which includes only stars that meet the photometric quality criteria described in the previous section and have measurable PMs. Panels\,(a) and (b) show the vector-point diagram (VPD) and the $m_{\rm F322W2}$ versus $m_{\rm F150W2}-m_{\rm F322W2}$ CMD, respectively. Since PMs are calculated relative to the cluster's overall motion, the distribution of cluster members in the VPD is centred at (0,0). A second, more sparse group of points is visible in the upper part of the VPD, representing background and foreground sources. Panel\,(c) shows the one-dimensional PM ($\mu_{\rm R}$, obtained by summing the PMs in the two directions in quadrature) plotted against $m_{\rm F322W2}$. $\omega$\,Cen members exhibit a narrow distribution in $\mu_{\rm R}$, mostly clustered below $\mu_{\rm R}<2$ mas yr$^{-1}$, while field objects extend towards higher $\mu_{\rm R}$. We defined a conservative PM selection to separate cluster members from field objects, indicated by a red line. Panels\,(d) and (e) show the VPD and $m_{\rm F322W2}$ versus $m_{\rm F150W2}-m_{\rm F322W2}$ CMD for the stars that passed the PM selection, while panels\,(f) and (g) show the same diagrams for stars that did not pass the PM selection. In panels\,(b), (e), and (g), the light-red and dark-red shaded regions indicate areas of the CMDs affected by mild and severe saturation, respectively. The PMs of stars in these regions should be treated with caution. In what follows, our analysis will focus exclusively on the sample of stars shown in panels\,(d) and (e).

\section{The JWST colour-magnitude diagram of $\omega$\,Cen}\label{Section4}

In the CMDs presented in Fig.\,\ref{CMDs}, \ref{CMDs2} and \ref{PMs}, the MS appears to be composed of two distinct components, which are clearly visible in the magnitude range from $m_{\rm F150W2} \sim$18 ($m_{\rm F322W2} \sim$17.5) down to $m_{\rm F150W2} \sim 20$ ($m_{\rm F322W2} \sim$19.5). At this point, the two sequences intersect and exchange positions. The sequence that is bluer above the intersection (and redder below) corresponds to the bMS population, while the sequence that is redder above the intersection (and bluer below) corresponds to the rMS population.

Note that we choose to maintain the names --- bMS and rMS --- as they were originally introduced based on the identification of these sequences using optical filters, where the bMS appears on the blue side and the rMS on the red side of the CMD \citep[see][]{2004ApJ...605L.125B,2009A&A...507.1393B}. Although in the low-MS, which is the focus of our study, the two sequences appear inverted in the JWST CMD (with the rMS on the blue side and the bMS on the red side), we retain this nomenclature for consistency with previous works.

\begin{centering} 
\begin{figure}
 \includegraphics[width=\columnwidth]{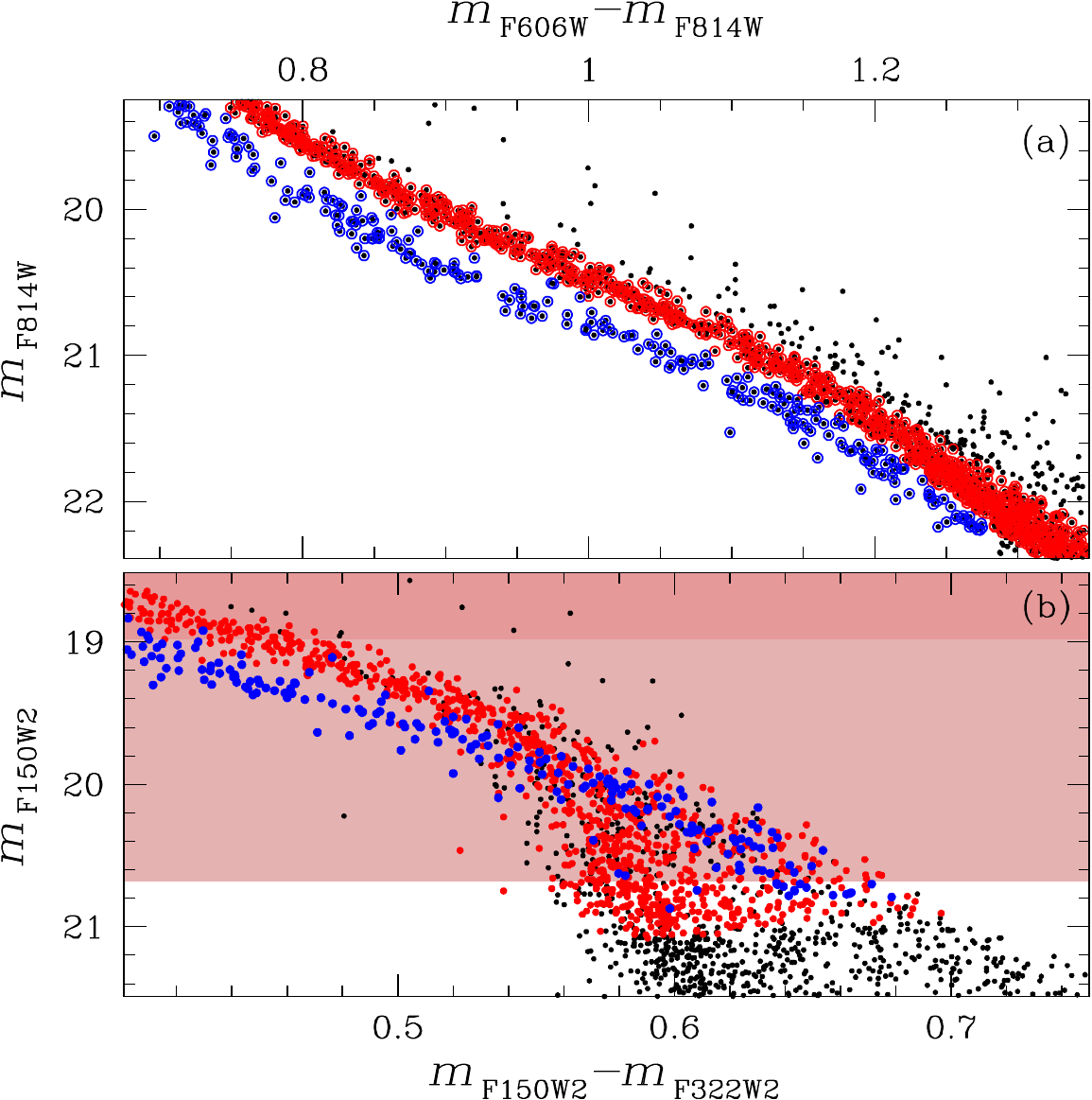}
 \caption{Crossing of the bMS and rMS stars in the JWST photometry-based CMD. 
 (a) In the HST $m_{\rm F814W}$ versus $m_{\rm F606W}-m_{\rm F814W}$ CMD, we isolated a sample of bMS (black points within blue circles) and rMS (black points within red circles) stars; the two sequences are clearly distinguishable. (b) The same samples highlighted in the JWST $m_{\rm F150W2}$ versus $m_{\rm F150W2}-m_{\rm F322W2}$ CMD, using the same colour code. The light-red and dark-red shaded regions highlight areas of the CMDs affected by mild and severe saturation, respectively.}
 \label{crossing} 
\end{figure} 
\end{centering}

To illustrate this inversion, in Fig.\,\ref{crossing}, we have isolated a sample of bMS and rMS stars from the CMD based on optical HST photometry and shown their position on the JWST CMD. As clearly shown, the bMS crosses the rMS around $m_{\rm F150W2} \sim 20$, becoming redder at fainter magnitudes. Below this crossing point, rMS stars span a broad colour range, but their distribution appears to favour bluer colours.

We note that, despite the onset of saturation at $m_{\rm F150W2} \sim 20.7$, the two main sequences remain clearly distinguishable in the JWST CMD, with a colour separation that appears well defined throughout the saturated regime. Although saturation introduces photometric discontinuities, it affects both sequences in a similar way and does not prevent a qualitative assessment of their morphology in this region.

Beyond this intersection ($m_{\rm F150W2}>20$ or $m_{\rm F322W2}>19.5$), the two main groups of stars remain distinguishable down to their respective terminations (see Fig.\,\ref{CMDs}, \ref{CMDs2} and \ref{PMs}). The rMS seems to terminate at $m_{\rm F150W2} \sim 24.75$ ($m_{\rm F322W2} \sim 24$), while the bMS ends $\sim0.5$ magnitude brighter ($m_{\rm F150W2} \sim 24.25$ or $m_{\rm F322W2} \sim 23.5$). To determine whether these terminations are real and not influenced by completeness effects, we conducted AS tests to assess the completeness of our catalogue, which will be discussed in the next Section.

\subsection{Chemical spread due to multiple populations}

The scatter in photometric colour is driven by the presence of mPOPs, and it encodes the chemical spread in $\omega$\,Cen. In order to interpret this information, we compared the observed CMD to theoretical isochrones. At this stage, we restricted our analysis to the overall range of $m_{\rm F150W2}-m_{\rm F322W2}$ colour in the CMD as a function of magnitude, and ignored all finer details of the underlying colour distribution. A likelihood-based analysis of star densities in the colour-magnitude space is deferred to a dedicated future study.

\begin{centering} 
\begin{figure*}
\centering
 \includegraphics[width=0.8\textwidth]{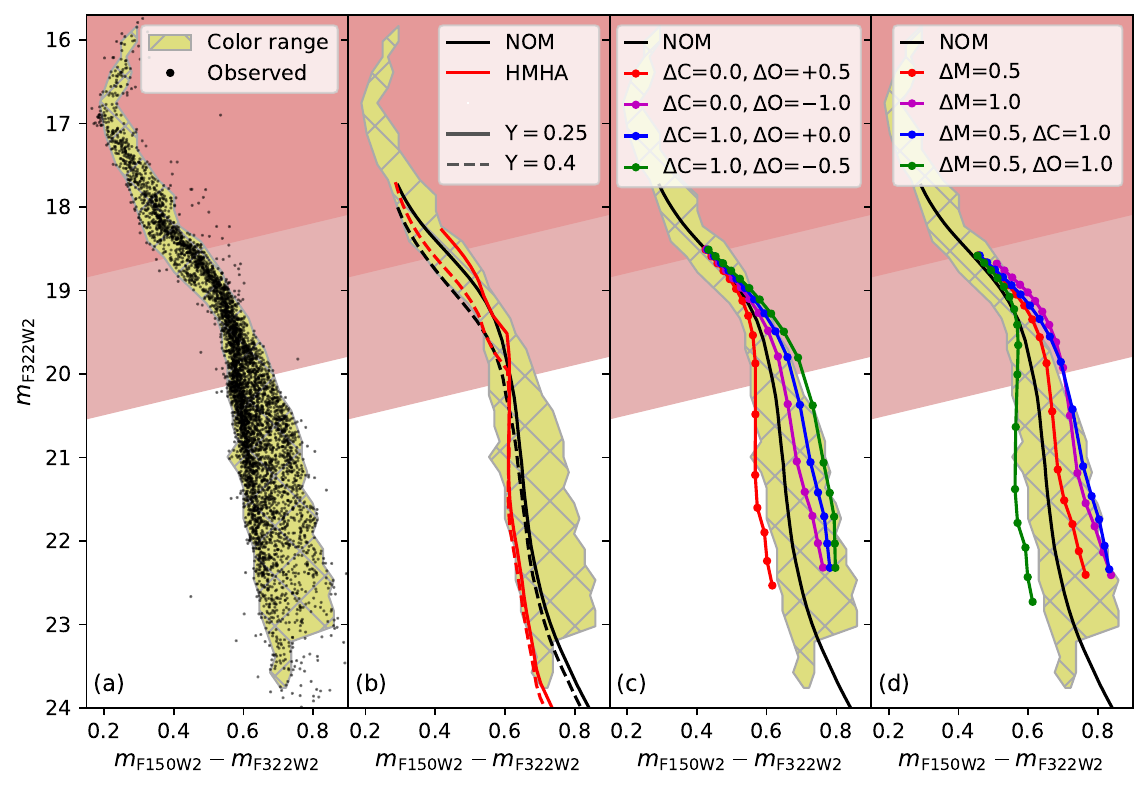}
 \caption{Comparison between the \textit{JWST} NIR CMD of $\omega$\,Cen presented in this study, and model isochrones. For clarity, we do not plot isochrones on the observed CMD directly. Instead, we compare the isochrones to the region of the colour-magnitude space occupied by the cluster, which is shaded in yellow in all panels. Panel (a) demonstrates how this region was chosen in relation to the observed CMD. Panel (b) over-plots two of the isochrones from \citet{2022ApJ...930...24G} with both their original helium mass fraction ($\mathrm{Y}=0.4$, dashed lines) and with modified evolutionary models that incorporate a near-solar helium abundance ($\mathrm{Y}=0.25$, solid lines). Panels (c) and (d) show the $\mathrm{Y}=0.25$ NOM isochrone as well as its lower-MS variations with various chemical offsets. In the legend, $\Delta \mathrm{C}$, $\Delta \mathrm{O}$ and $\Delta \mathrm{M}$ refers to the offsets in $[\mathrm{C/H}]$, $[\mathrm{O/H}]$ and $[\mathrm{M/H}]$ from the chemical composition adopted for NOM in \citet{2022ApJ...930...24G}, where $\mathrm{M}$ only refers to metals heavier than oxygen. The light-red and dark-red shaded regions highlight areas of the CMDs affected by mild and severe saturation, respectively.}
 \label{fig:chem} 
\end{figure*} 
\end{centering}

At each magnitude, we defined the observed colour range as the interval between the bluest and reddest colour, for which the star density in the colour-magnitude space exceeds a certain threshold that was adjusted ``by hand'' to capture the bulk of the colour distribution and exclude outliers. The calculated range is highlighted with yellow shading in panel (a) of Fig.\,\ref{fig:chem}. As the basis of our comparison, we adopted two isochrones from \citet{2022ApJ...930...24G} that the authors refer to as ``nominal'' (NOM) and ``High Metal High Alpha'' (HMHA). Both isochrones were calculated for the helium-enriched sub-population of the cluster with $\mathrm{Y}=0.4$ \citep{2012AJ....144....5K} that was first recognised in \citet{2004ApJ...605L.125B}. The NOM isochrone uses the modal light element ($\mathrm{C}$, $\mathrm{N}$ and $\mathrm{O}$) abundances and metallicity ($[\mathrm{Fe/H}]=-1.7$) from the spectroscopic observations of giant members in \citet{spectroscopic_abundances}. The HMHA isochrone was derived from NOM, but with metallicity and $\alpha$-enhancement (including $\mathrm{O}$) tuned to match the ridgelines of the NIR and optical CMDs from \textit{HST} photometry. In particular, the HMHA model adopts $[\mathrm{M/H}] = -1.4$, $[\mathrm{\alpha/M}] = +0.6$, as detailed in \citet{2022ApJ...930...24G}. The overall range of $[\mathrm{O/H}]$ covered by these two isochrones spans $1\ \mathrm{dex}$, and overlaps with $\sim80\%$ of the spectroscopically inferred distribution\footnote{The $[\mathrm{O/H}]=[\mathrm{O/M}]+[\mathrm{M/H}]$ abundance ratios adopted in NOM and HMHA isochrones are $-1.8$ and $-0.8$, respectively (Tables\,1 and 2 of \citealt{2022ApJ...930...24G}). These values correspond to the $\sim12$th and $\sim92$nd percentiles of the spectroscopic distribution of $[\mathrm{O/H}]$ (upper right panel of Fig.\,3 of \citealt{spectroscopic_abundances}).}.

In order to explore the full range of helium mass fraction in $\omega$\,Cen, we calculated alternative sets of evolutionary models for both NOM and HMHA with $\mathrm{Y}=0.25$ using the \texttt{MESA} code \citep{MESA,MESA_2,MESA_3,MESA_4,MESA_5}, and following the method from \citet{SANDee}. We note that the modified versions of the isochrones retain the original atmosphere models with $\mathrm{Y}=0.4$. Therefore, they have inconsistent helium mass fractions between stellar atmospheres and interiors. We anticipate the error due to this inconsistency to be insignificant since the shape of the lower MS is insensitive to $\mathrm{Y}$ \citep{paper_IV}, and the effect of $\mathrm{Y}$ on the upper MS largely originates in the stellar interiors rather than atmospheres \citep{salaris_textbook}. We transformed the isochrones to the observed plane using the same parameters and procedure as in \citet{2022ApJ...930...24G}, but with a lower interstellar reddening ($E(B-V)=0.12$, which is more consistent with other measurements in the literature, \citealt{1996AJ....112.1487H,2005ApJ...634L..69C}) and a more up-to-date reddening law from \citet{extinction}.

All four model isochrones (the original and modified versions of NOM and HMHA) are compared to the observed colour spread in panel (b) of Fig.\,\ref{fig:chem}.The isochrones are also provided in Tables\,\ref{tab:NOM25}, \ref{tab:NOM40}, \ref{tab:HMHA25}, and \ref{tab:HMHA40} in Appendix\,\ref{Appendix:B}. Both the isochrones and the observed CMD display a clear change of slope near $m_{\rm F322W2}\approx19.5$. This feature is known as the \textit{MS knee} \citep{MS_knee,MS_knee_3}, and it is caused by the reduction in the adiabatic gradient due to dissociation of molecular hydrogen \citep{SCHM,MS_inflection_3}. Above the knee, the effect of atmospheric chemistry on the shape of the isochrone is subtle, since high effective temperatures ($T_\mathrm{eff}\gtrsim4500\ \mathrm{K}$) suppress the abundances of key infrared absorbers such as $\mathrm{H_2O}$. The small residual effect is driven by two primary factors. First, the mean opacity of the atmosphere reduces $T_\mathrm{eff}$ at higher metallicities and makes the star redder. This effect is captured in the \textit{atmosphere-interior coupling} scheme adopted in \citet{2022ApJ...930...24G}. The second factor is related to the hydrogen anion ($\mathrm{H}^-$) absorption that has a minimum near $\qty{1.6}{\micro\meter}$, producing the characteristic ``$\mathrm{H}^-$ bump'' in the spectrum \citep{Hmin_1,Hmin_2}. The bump becomes more pronounced at higher metallicities, making the $m_{\rm F150W2}-m_{\rm F322W2}$ colour of the star bluer.

The $m_{\rm F150W2}-m_{\rm F322W2}$ CMD therefore provides a unique opportunity to investigate stellar interiors, since these two already small colour-metallicity dependencies due to stellar atmospheres suppress one another. For this reason, the colour of stars above the MS knee is predominantly determined by the stellar interior, and not the atmosphere. In particular, the observed colour of the star is expected to be largely determined by the mean molecular weight of the interior, which is most sensitive to the helium mass fraction, $\mathrm{Y}$. Panel (b) of Fig.\,\ref{fig:chem} clearly shows that $\mathrm{Y}$ has the dominant impact on the $m_{\rm F150W2}-m_{\rm F322W2}$ colour, despite the NOM and HMHA isochrones having vastly different chemistries in the atmospheres. The scatter in $\mathrm{Y}$ of $\sim 0.15$ is required to fully capture the observed width of the CMD above the MS knee.

The scatter in the CMD colour is noticeably wider below the MS knee due to the increasing effect of atmospheric chemistry including, most notably, infrared absorption bands of $\mathrm{H_2O}$ that are regulated directly by $[\mathrm{O/Fe}]$, and indirectly by $[\mathrm{C/Fe}]$ through the carbon-oxygen chemical reaction network of the atmosphere \citep{CO_ratio}. Both $[\mathrm{C/Fe}]$ and $[\mathrm{O/Fe}]$ are expected to display large member-to-member variations in GCs due to mPOPs. To determine if these variations alone are sufficient to explain the observed CMD below the MS knee, we adopted the $T_\mathrm{eff}-\log(g)$ and $T_\mathrm{eff}$-luminosity relationships from the $\mathrm{Y}=0.25$ NOM isochrone, and calculated additional model atmospheres for various values of $[\mathrm{C/Fe}]/[\mathrm{O/Fe}]$ at $3500\ \mathrm{K}\leq T_\mathrm{eff}\leq5000\ \mathrm{K}$ in $100\ \mathrm{K}$ steps. The new models were calculated using the \texttt{BasicATLAS}/\texttt{ATLAS-9}/\texttt{SYNTHE} setup \citep{BasicATLAS,ATLAS5,ATLAS9_1,ATLAS9_2,SYNTHE}.

The upper temperature limit for the new models was chosen to approximately match the severe saturation limit of our photometry (also shown in Fig.\,\ref{fig:chem} with dark-red shading). The lower temperature limit was chosen out of practical considerations, as \texttt{ATLAS} atmospheres are expected to become increasingly unreliable at $T_\mathrm{eff}\lesssim 3500\ \mathrm{K}$ \citep{ATLAS_eval}. Due to the low metallicity of $\omega$\,Cen, the effective temperature of $3500\ \mathrm{K}$ corresponds to comparatively low stellar masses, varying between $0.10\ M_\odot$ and $0.15\ M_\odot$ among the four isochrones introduced above. We therefore expect the temperature range of new model atmospheres to be sufficiently wide to accommodate nearly all of our data.

For each set of chemical offsets from the composition of the NOM isochrone, we only calculated the updated synthetic spectra, and ignored the effect of altered chemistry on the atmospheric structures, atmosphere-interior coupling and stellar evolution. While these effects are expected to be subdominant on the lower MS, they may not be insignificant. For this reason, this analysis must be considered preliminary and taken with caution.

The synthetic colours and magnitudes of the new models are plotted alongside the NOM $\mathrm{Y}=0.25$ isochrone in panel (c) of Fig.\,\ref{fig:chem}. The figure shows that a spread in $[\mathrm{C/Fe}]$/$[\mathrm{O/Fe}]$ alone is sufficient to explain the full width of the observed CMD; however, the red tail of the colour distribution requires a significant fraction of the stars to have simultaneously very low oxygen and very high carbon abundances, which is unlikely due to the strong carbon-oxygen correlation that is expected in GCs in general (see review in \citealt{mPOPs_review}), and has been spectroscopically confirmed in $\omega$\,Cen in particular \citep{spectroscopic_abundances}.

\begin{centering} 
\begin{figure}
 \includegraphics[width=\columnwidth]{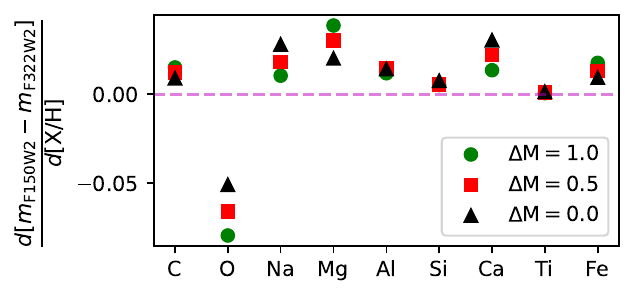}
 \caption{Effect of offsets in the abundances of selected metals from the chemical composition of the NOM isochrone (panels (b)-(d) of Fig.\,\ref{fig:chem}) on the observed $m_{\rm F150W2}-m_{\rm F322W2}$ colour of the star at $T_\mathrm{eff}=4000\ \mathrm{K}$. The effects are shown at three different metallicity offsets, where $\mathrm{M}$ only includes metals heavier than oxygen. Positive values indicate that the star gets redder when the abundance of the element is increased. The $0$ (no-effect) horizontal line is shown for reference.} 
 \label{fig:variations} 
\end{figure} 
\end{centering}

Alternatively, the red tail of the colour distribution may be due to member-to-member variations in metallicity that have been spectroscopically confirmed in this GC \citep{2011ApJ...731...64M}. Panel (d) of Fig.\,\ref{fig:chem} demonstrates that the red tail of the CMD can be approximated either by models with large but plausible metallicity offsets from NOM, or by more modest metallicity offsets in combination with adjusted light element abundances. The effect of individual metals on the $m_{\rm F150W2}-m_{\rm F322W2}$ colour of the star at $T_\mathrm{eff}=4000\ \mathrm{K}$ is shown in Fig.\,\ref{fig:variations}. With the exception of oxygen that directly influences infrared absorption features in the spectrum, the most important metal abundances are those of key electron donors (e.g., $\mathrm{Mg}$, $\mathrm{Ca}$, $\mathrm{Na}$) that alter the slope of the stellar continuum.

As seen in both panels (c) and (d) of Fig.\,\ref{fig:chem}, appropriately chosen isochrones can be used to approximate either tail of the distribution across most of the observed lower MS. We therefore conclude that our observations of $\omega$\,Cen do not suggest that the chemical spread in the cluster varies with stellar mass.

\section{Artificial stars}\label{Section5}

\begin{centering} 
\begin{figure}
 \includegraphics[width=\columnwidth]{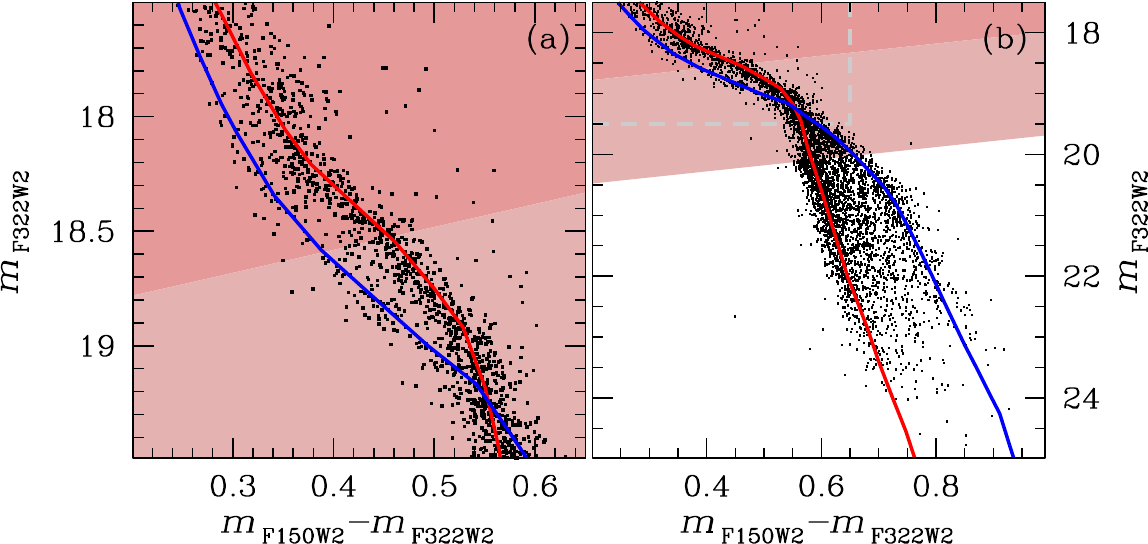}
 \caption{$m_{\rm F322W2}$ versus $m_{\rm F150W2}-m_{\rm F322W2}$ CMDs of our selected sample of stars, focusing on the upper-MS (panel a) and the entire MS (panel b), with the two fiducials used to generate the ASs. The grey dashed rectangle in panel b highlights the CMD region displayed in panel a. The blue fiducial corresponds to the bMS, while the red fiducial corresponds to the rMS. In the upper part of the MS, the bMS occupies the blue side of the CMD, while the rMS is on the red side. Below the intersection point, the two sequences switch positions. The light-red and dark-red shaded regions highlight areas of the CMDs affected by mild and severe saturation, respectively.}
 \label{FID} 
\end{figure} 
\end{centering}

We used ASs to estimate the completeness of our catalogue for each of the two sequences. We first evaluated the completeness of the JWST data by generating ASs uniformly distributed across the JWST FOV. This allowed us to evaluate the completeness level for the CMD that includes all stars within the JWST field (see Fig.\,\ref{CMDs}). The magnitudes in F322W2 were uniformly distributed within the range $18 < m_{\rm F322W2} < 28.5$, while the corresponding F150W2 magnitudes were assigned based on fiducial lines manually defined in the $m_{\rm F322W2}$ versus $m_{\rm F150W2}-m_{\rm F322W2}$ CMD. One fiducial line was drawn for each of the two sequences, as shown in Fig.\,\ref{FID}. We generated 50,000 ASs for each sequence, for a total of 100,000 ASs. The ASs were generated, detected, and measured using \texttt{KS2}, following the same procedures used for the real stars. An AS was considered recovered if the difference between the input and output positions was less than 1 pixel, the difference between the input and output magnitudes was within 0.75 mag (equivalent to $\sim$2.5log2), and it passed the same selection criteria applied to the real stars.

\begin{centering} 
\begin{figure*}
\centering
 \includegraphics[width=0.8\textwidth]{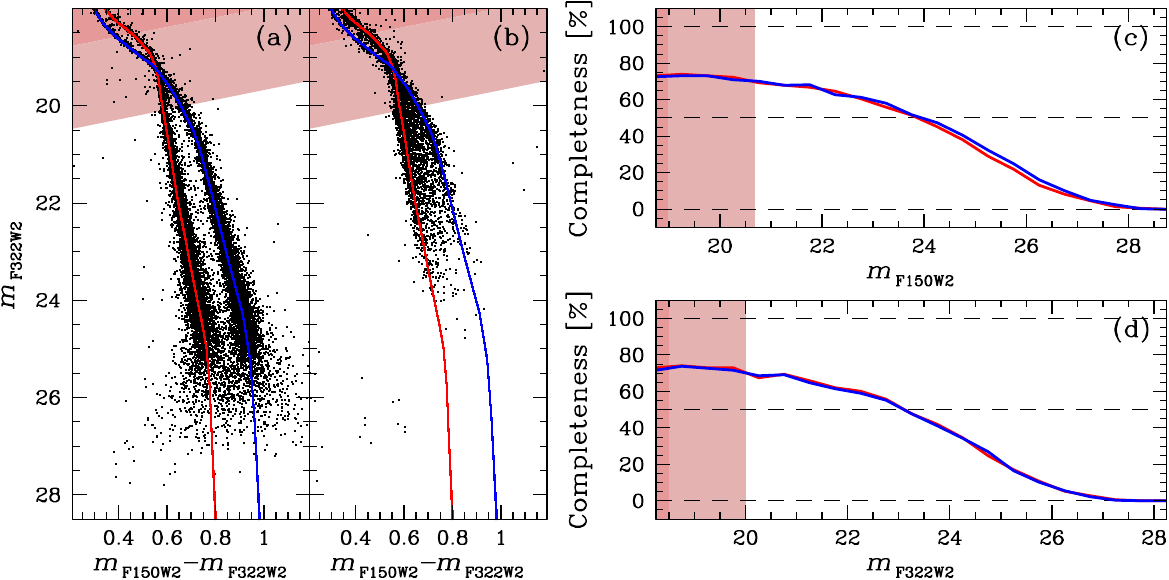}
 \caption{(a) Injected ASs for the two sequences (blue and red points) in the $m_{\rm F150W2}$ versus $m_{\rm F150W2}-m_{\rm F322W2}$ CMD. Recovered ASs are represented in black. (b) Same as panel (a), but the black points now represent the real stars. (c)-(d) Completeness as a function of the $m_{\rm F150W2}$ and $m_{\rm F322W2}$ magnitude, respectively. The completeness for each sequence is represented using the same colour scheme as in panel\,(a). Light-red and dark-red shaded areas qualitatively mark magnitude ranges where MS stars are mildly or severely saturated in at least one filter.}
 \label{ASs} 
\end{figure*} 
\end{centering}

Panel\,(a) of Fig.\,\ref{ASs} shows the injected ASs for the two sequences in the $m_{\rm F322W2}$ versus $m_{\rm F150W2}-m_{\rm F322W2}$ CMD, with injected stars in red and blue, and recovered stars in black. For comparison, panel\,(b) presents the same CMD for the real stars, with the two fiducial lines overplotted, using the same scale as in panel\,(a). As clearly visible from panel\,(a), the recovered ASs exhibit a slight asymmetry in their colour distribution, with redward tails extending from both sequences. This behaviour stems from a known effect of crowding in dense stellar fields. In particular, ASs that fall near bright real stars can be recovered as slightly brighter and redder. This systematic effect produces a characteristic skewness in the colour distribution, manifesting as redward tails.

The completeness, defined as the ratio of recovered stars to the total number of injected stars, is shown as a function of $m_{\rm F150W2}$ and $m_{\rm F322W2}$ magnitudes in panels\,(b) and (c), for both sequences. In these panels, the completeness for each sequence is represented using the same colour scheme as in panel\,(a). In this and the other figures of this section, light-red and dark-red shaded areas qualitatively mark magnitude ranges where MS stars are mildly or severely saturated in at least one filter. Completeness estimates in these magnitude ranges should be interpreted with caution.

From panel\,(a), it is evident that the recovered ASs extend well beyond the termination points observed in the real stars. The recovered ASs reach the same depth for both sequences, showing no signs of the differing termination magnitudes observed in the real stars. The completeness levels for the two sequences, shown in panels\,(b) and (c), are nearly identical. At the observed termination magnitudes of the sequences in the real stars ($m_{\rm F322W2}\sim 24$ for the rMS and $m_{\rm F322W2}\sim 23.5$ for the bMS), the completeness remains relatively high, around $\sim 40\%$.

These results confirm that the observed terminations in the two sequences are real and not influenced by completeness limitations. The absence of a significant difference in completeness levels between the two sequences further supports the conclusion that the earlier termination of the bMS is an intrinsic feature rather than a consequence of completeness effects.

\begin{centering} 
\begin{figure}
 \includegraphics[width=\columnwidth]{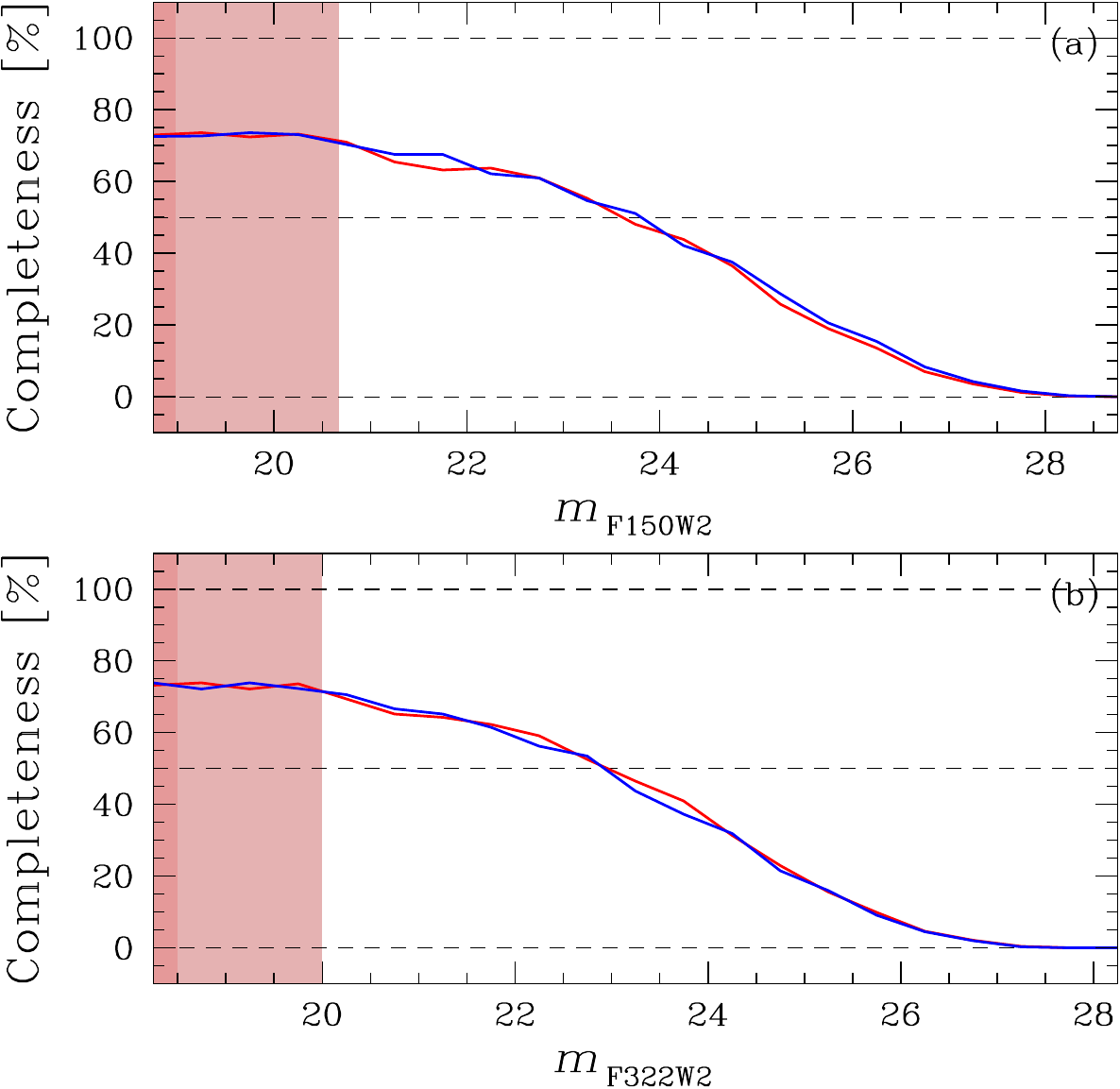}
 \caption{Completeness as a function of $m_{\rm F150W2}$ (panel a) and $m_{\rm F322W2}$ (panel b) magnitudes for the two sequences in the overlapping region of the HST and JWST fields.} 
 \label{ASs2} 
\end{figure} 
\end{centering}

To evaluate the completeness level for the CMD that includes all stars in the overlapping FOV of both HST and JWST (see Fig.\,\ref{CMDs2}), we generated an additional set of 100,000 ASs (50,000 for each sequence), uniformly distributed across the common area. The resulting completeness levels are shown in Fig.\,\ref{ASs2} as a function of the $m_{\rm F150W2}$ (panel a) and $m_{\rm F322W2}$ (panel b) magnitudes. The completeness values are nearly identical to those obtained when considering the entire JWST field.

To incorporate PM selection into our completeness evaluation and assess the completeness for the stars shown in panel\,(e) of Fig.\,\ref{PMs}, we performed the ASs procedure for the HST data as well. We used the same positions and F150W2 and F322W2 magnitudes defined above and determined the corresponding F606W and F814W magnitudes using fiducial lines manually defined in the $m_{\rm F322W2}$ versus $m_{\rm F814W}-m_{\rm F322W2}$ and $m_{\rm F322W2}$ versus $m_{\rm F606W}-m_{\rm F322W}$ CMDs. We ran the ASs stars separately for the JWST and HST datasets and then assessed the displacement of the stars between the two epochs.

Since ASs are generated with identical positions in both epochs, any observed displacement is solely due to spurious offsets caused by noise (e.g., uncertainties in PSF modelling, coordinate transformations, cosmic ray hits, detector cosmetics, etc.), which introduces slight shifts in the recovered positions across epochs. To be considered recovered, an AS must also satisfy the same PM selection criteria applied to real stars, as shown in panel\,(d) of Fig.\,\ref{PMs}. By applying this PM selection to ASs, we estimated the fraction of real stars excluded due to these positional offsets, allowing us to incorporate this effect into the overall completeness calculation.

\begin{centering} 
\begin{figure}
 \includegraphics[width=\columnwidth]{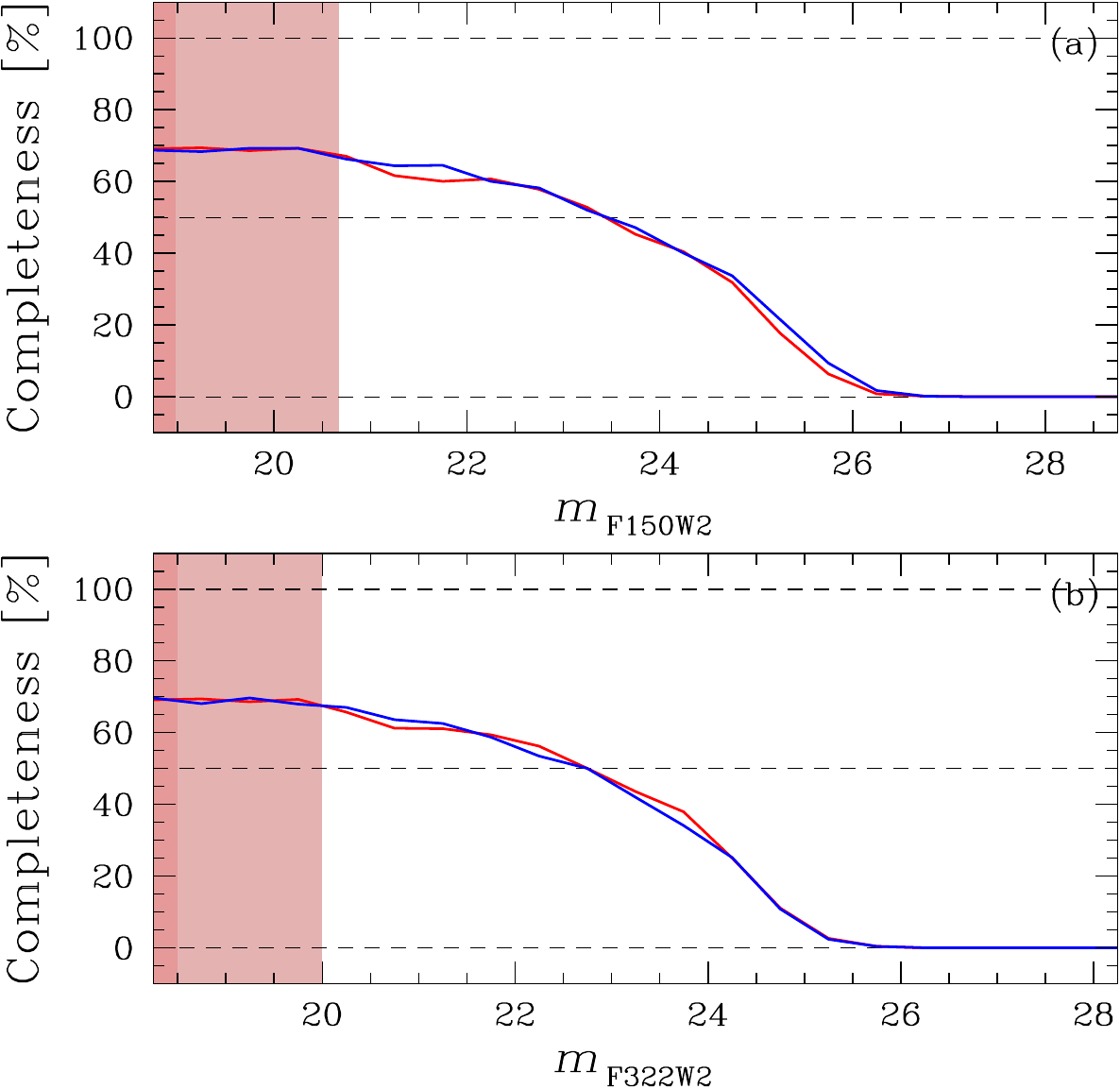}
 \caption{Same as Fig.\,\ref{ASs2} but including the PMs selection.} 
 \label{ASs3} 
\end{figure} 
\end{centering}

The completeness obtained after applying the PM selection is shown in Fig.\,\ref{ASs3}. As expected, the completeness of both sequences is lower and declines more rapidly toward fainter magnitudes compared to Fig.\,\ref{ASs2}, reaching zero at a brighter magnitude. However, at the termination points of the sequences for real stars ($m_{\rm F322W2} \sim 24$ for the rMS and $m_{\rm F322W2} \sim 23.5$ for the bMS; see panel\,(d) of Fig.\,\ref{PMs}), the completeness remains relatively high, around $\sim$30–40\%, confirming the reliability of the sequences termination. Given the consistent completeness between the two sequences, we will adopt the completeness values obtained for the rMS in the following analysis.

\section{Luminosity and mass function}\label{Section6}
In this section, we present the LF and MF of $\omega$\,Cen and its sequences using JWST photometry.

\begin{table*}
\caption{Best-fit parameters of the LF models considered in this study. Population 1 and 2 refer to the isochrones in panel (b) of Fig.\,\ref{fig:chem}, whose MLR have been used. ``25'' and ``40'' at the end of the population name refer to the $\mathrm{Y}=0.25$ and $\mathrm{Y}=0.4$ variants of the isochrones. GoF is the goodness-of-fit as defined in the text. Smaller GoF values imply better fits. See Section\,\ref{Section6.1} for the definition of the parameters.}
\centering
\begin{tabular}{lll|cccccc}
\hline
\multirow{2}{*}{\textbf{Band}} & \multirow{2}{*}{\textbf{Population 1}} & \multirow{2}{*}{\textbf{Population 2}}   &
\multirow{2}{*}{$\alpha_\mathrm{h}$} & \multirow{2}{*}{$\alpha_\mathrm{l}$} & \multirow{2}{*}{$M_\mathrm{pb}/M_\odot$} & \multirow{2}{*}{$\alpha_2$} & \multirow{2}{*}{$\gamma$} & \multirow{2}{*}{GoF}\\
 &  &   &
 &  &  &  & & \\\hline
\hline\multicolumn{9}{l}{\textit{Single population with single-component power-law MF}} \\\hline
\multirow{4}{*}{F150W2} & NOM25 & N/A & $0.77^{+0.03}_{-0.03}$ & N/A & N/A & N/A & N/A & 1.66 \\
 & NOM40 & N/A & $0.53^{+0.03}_{-0.03}$ & N/A & N/A & N/A & N/A & 1.84 \\
 & HMHA25 & N/A & $0.57^{+0.03}_{-0.03}$ & N/A & N/A & N/A & N/A & 1.90 \\
 & HMHA40 & N/A & $0.37^{+0.03}_{-0.03}$ & N/A & N/A & N/A & N/A & 2.56 \\
\hline
\multirow{4}{*}{F322W2} & NOM25 & N/A & $0.72^{+0.03}_{-0.03}$ & N/A & N/A & N/A & N/A & 2.03 \\
 & NOM40 & N/A & $0.46^{+0.03}_{-0.03}$ & N/A & N/A & N/A & N/A & 1.93 \\
 & HMHA25 & N/A & $0.46^{+0.03}_{-0.03}$ & N/A & N/A & N/A & N/A & 2.21 \\
 & HMHA40 & N/A & $0.21^{+0.04}_{-0.04}$ & N/A & N/A & N/A & N/A & 2.17 \\
\hline\multicolumn{9}{l}{\textit{Single population with two-component broken power-law MF}} \\\hline
\multirow{4}{*}{F150W2} & NOM25 & N/A & $1.05^{+0.09}_{-0.07}$ & $0.13^{+0.15}_{-0.22}$ & $0.19^{+0.03}_{-0.02}$ & N/A & N/A & 0.69 \\
 & NOM40 & N/A & $1.64^{+0.33}_{-0.25}$ & $0.34^{+0.05}_{-0.05}$ & $0.31^{+0.02}_{-0.02}$ & N/A & N/A & 1.09 \\
 & HMHA25 & N/A & $0.94^{+0.11}_{-0.10}$ & $-0.06^{+0.13}_{-0.41}$ & $0.21^{+0.03}_{-0.03}$ & N/A & N/A & 0.90 \\
 & HMHA40 & N/A & $2.25^{+0.35}_{-0.29}$ & $0.08^{+0.05}_{-0.05}$ & $0.32^{+0.01}_{-0.01}$ & N/A & N/A & 1.10 \\
\hline
\multirow{4}{*}{F322W2} & NOM25 & N/A & $1.20^{+0.13}_{-0.12}$ & $0.17^{+0.11}_{-0.27}$ & $0.21^{+0.02}_{-0.03}$ & N/A & N/A & 0.84 \\
 & NOM40 & N/A & $2.35^{+0.81}_{-0.48}$ & $0.29^{+0.05}_{-0.06}$ & $0.32^{+0.01}_{-0.02}$ & N/A & N/A & 1.29 \\
 & HMHA25 & N/A & $1.17^{+0.13}_{-0.17}$ & $-0.04^{+0.08}_{-0.18}$ & $0.25^{+0.01}_{-0.03}$ & N/A & N/A & 0.88 \\
 & HMHA40 & N/A & $1.85^{+0.37}_{-0.27}$ & $-0.07^{+0.05}_{-0.06}$ & $0.29^{+0.01}_{-0.01}$ & N/A & N/A & 1.34 \\
\hline\multicolumn{9}{l}{\textit{Mixture of two populations with identical single-component power-law MFs}} \\\hline
\multirow{5}{*}{F150W2} & NOM25 & NOM40 & $0.69^{+0.04}_{-0.04}$ & N/A & N/A & N/A & $0.68^{+0.13}_{-0.14}$ & 1.61 \\
 & NOM25 & HMHA25 & $0.74^{+0.03}_{-0.04}$ & N/A & N/A & N/A & $0.89^{+0.05}_{-0.15}$ & 1.68 \\
 & NOM25 & HMHA40 & $0.74^{+0.03}_{-0.04}$ & N/A & N/A & N/A & $0.94^{+0.03}_{-0.08}$ & 1.68 \\
 & HMHA25 & NOM40 & $0.55^{+0.03}_{-0.03}$ & N/A & N/A & N/A & $0.41^{+0.15}_{-0.13}$ & 1.81 \\
 & HMHA25 & HMHA40 & $0.55^{+0.03}_{-0.04}$ & N/A & N/A & N/A & $0.88^{+0.06}_{-0.11}$ & 1.92 \\
\hline
\multirow{5}{*}{F322W2} & NOM25 & NOM40 & $0.64^{+0.05}_{-0.05}$ & N/A & N/A & N/A & $0.66^{+0.14}_{-0.14}$ & 1.91 \\
 & NOM25 & HMHA25 & $0.70^{+0.04}_{-0.05}$ & N/A & N/A & N/A & $0.92^{+0.04}_{-0.13}$ & 2.03 \\
 & NOM25 & HMHA40 & $0.68^{+0.04}_{-0.06}$ & N/A & N/A & N/A & $0.93^{+0.03}_{-0.11}$ & 2.01 \\
 & HMHA25 & NOM40 & $0.46^{+0.03}_{-0.04}$ & N/A & N/A & N/A & $0.38^{+0.13}_{-0.12}$ & 1.97 \\
 & HMHA25 & HMHA40 & $0.36^{+0.05}_{-0.05}$ & N/A & N/A & N/A & $0.55^{+0.15}_{-0.13}$ & 2.17 \\
\hline\multicolumn{9}{l}{\textit{Mixture of two populations with identical two-component broken power-law MFs}} \\\hline
\multirow{5}{*}{F150W2} & NOM25 & NOM40 & $0.98^{+0.10}_{-0.08}$ & $0.03^{+0.18}_{-0.36}$ & $0.17^{+0.03}_{-0.02}$ & N/A & $0.72^{+0.12}_{-0.13}$ & 0.66 \\
 & NOM25 & HMHA25 & $1.02^{+0.09}_{-0.08}$ & $0.05^{+0.15}_{-0.28}$ & $0.19^{+0.03}_{-0.02}$ & N/A & $0.68^{+0.14}_{-0.18}$ & 0.66 \\
 & NOM25 & HMHA40 & $1.03^{+0.11}_{-0.09}$ & $0.09^{+0.14}_{-0.31}$ & $0.19^{+0.03}_{-0.03}$ & N/A & $0.80^{+0.09}_{-0.11}$ & 0.71 \\
 & HMHA25 & NOM40 & $0.83^{+0.17}_{-0.07}$ & $-0.36^{+0.30}_{-0.50}$ & $0.16^{+0.05}_{-0.02}$ & N/A & $0.67^{+0.14}_{-0.16}$ & 0.89 \\
 & HMHA25 & HMHA40 & $0.96^{+0.23}_{-0.13}$ & $-0.03^{+0.12}_{-0.46}$ & $0.22^{+0.04}_{-0.04}$ & N/A & $0.83^{+0.08}_{-0.17}$ & 0.99 \\
\hline
\multirow{5}{*}{F322W2} & NOM25 & NOM40 & $1.12^{+0.16}_{-0.15}$ & $0.12^{+0.12}_{-0.41}$ & $0.20^{+0.02}_{-0.03}$ & N/A & $0.81^{+0.09}_{-0.14}$ & 0.80 \\
 & NOM25 & HMHA25 & $1.19^{+0.13}_{-0.14}$ & $0.07^{+0.12}_{-0.25}$ & $0.22^{+0.02}_{-0.03}$ & N/A & $0.69^{+0.14}_{-0.18}$ & 0.82 \\
 & NOM25 & HMHA40 & $1.20^{+0.14}_{-0.18}$ & $0.11^{+0.11}_{-0.39}$ & $0.23^{+0.02}_{-0.04}$ & N/A & $0.81^{+0.09}_{-0.13}$ & 0.87 \\
 & HMHA25 & NOM40 & $0.98^{+0.25}_{-0.17}$ & $-0.14^{+0.15}_{-0.69}$ & $0.22^{+0.03}_{-0.05}$ & N/A & $0.71^{+0.13}_{-0.16}$ & 1.06 \\
 & HMHA25 & HMHA40 & $1.21^{+0.26}_{-0.29}$ & $-0.05^{+0.08}_{-0.54}$ & $0.26^{+0.02}_{-0.06}$ & N/A & $0.69^{+0.13}_{-0.16}$ & 0.99 \\
\hline\multicolumn{9}{l}{\textit{Mixture of two populations with distinct single-component power-law MFs}} \\\hline
\multirow{5}{*}{F150W2} & NOM25 & NOM40 & $0.79^{+0.08}_{-0.07}$ & N/A & N/A & $0.42^{+0.18}_{-2.50}$ & $0.59^{+0.19}_{-0.15}$ & 1.61 \\
 & NOM25 & HMHA25 & $0.78^{+0.04}_{-0.03}$ & N/A & N/A & $-2.55^{+1.67}_{-1.19}$ & $0.87^{+0.06}_{-0.19}$ & 1.69 \\
 & NOM25 & HMHA40 & $0.78^{+0.04}_{-0.03}$ & N/A & N/A & $-2.08^{+1.32}_{-1.41}$ & $0.78^{+0.10}_{-0.21}$ & 1.66 \\
 & HMHA25 & NOM40 & $0.54^{+0.12}_{-2.05}$ & N/A & N/A & $0.55^{+0.09}_{-0.10}$ & $0.41^{+0.16}_{-0.18}$ & 1.81 \\
 & HMHA25 & HMHA40 & $0.58^{+0.04}_{-0.03}$ & N/A & N/A & $-2.90^{+1.75}_{-1.02}$ & $0.85^{+0.07}_{-0.22}$ & 1.98 \\
\hline
\multirow{5}{*}{F322W2} & NOM25 & NOM40 & $0.76^{+0.09}_{-0.08}$ & N/A & N/A & $0.35^{+0.16}_{-2.06}$ & $0.52^{+0.21}_{-0.14}$ & 1.83 \\
 & NOM25 & HMHA25 & $0.73^{+0.05}_{-0.04}$ & N/A & N/A & $-2.57^{+1.60}_{-1.18}$ & $0.85^{+0.07}_{-0.22}$ & 2.05 \\
 & NOM25 & HMHA40 & $0.75^{+0.07}_{-0.05}$ & N/A & N/A & $-0.11^{+0.26}_{-2.34}$ & $0.60^{+0.19}_{-0.17}$ & 2.00 \\
 & HMHA25 & NOM40 & $0.43^{+0.13}_{-0.19}$ & N/A & N/A & $0.48^{+0.09}_{-0.10}$ & $0.38^{+0.15}_{-0.14}$ & 1.95 \\
 & HMHA25 & HMHA40 & $0.50^{+0.09}_{-0.09}$ & N/A & N/A & $0.15^{+0.12}_{-0.16}$ & $0.44^{+0.17}_{-0.12}$ & 2.11 \\
\hline
\end{tabular}
\label{tab:MF_params}
\end{table*}

\subsection{Combined luminosity and mass functions}\label{Section6.1}

We derived the combined (i.e., without matching individual stars to sub-populations) LF and MF of $\omega$\,Cen by fitting a forward model to the observed $m_\mathrm{F150W2}$ and $m_\mathrm{F322W2}$ magnitudes of confirmed members. The likelihood of compatibility between the forward model and a set of photometric measurements with uncertainties, $\{m_i,
\sigma_i\}$, is given by the following equation:

\begin{equation}
    \mathcal{L}=\prod_i \int_{m_\mathrm{min}}^\infty \phi(m_i)\ \mathrm{comp}(m_i)\ Z(x;m_i,\sigma_i)\ dx
    \label{eq:likelihood}
\end{equation}

\noindent where $\phi(m)$ is the LF (i.e., the likelihood of observing a member with magnitude $m$), $\mathrm{comp}(m)$ is the photometric completeness at magnitude $m$, and $Z(x;m,\sigma)$ is the probability of measuring magnitude $x$, given the true magnitude $m$ and the photometric error $\sigma$. We took $Z(x;m,\sigma)$ to be a normal distribution with the mean of $m$ and the standard deviation of $\sigma$. The brightest considered magnitude, $m_\mathrm{min}$, was set to the saturation limit of the instrument, which we conservatively set to $m_\mathrm{min}=19$ for both F150W2 and F322W2.

We considered $5$ different families of LF models ($\phi(m)$) listed below:

\begin{enumerate}
    \item \textit{Single population with single-component power-law MF}. To construct this model, we used the mass-luminosity relationship (MLR) from only one of the theoretical isochrones shown in panel (b) of Fig.\,\ref{fig:chem}. We took the MF to obey a single-component power-law distribution as follows:

\begin{equation}
\xi(M)\propto M^{-\alpha_\mathrm{h}}
\label{eq:unbroken_MF}
\end{equation}

    where $\xi(M)$ is the MF (i.e., the likelihood of observing a member with mass $M$), and $\alpha_\mathrm{h}$ is the slope of the power law. $\xi(M)$ was converted to $\phi(m)$ using the MLR of the adopted isochrone as follows:

\begin{equation}
\phi(m)\propto \xi(M(m)) \left| \frac{dM(m)}{dm}\right|
\label{eq:MF_to_LF}
\end{equation}

    where $M(m)$ is the MLR, and the derivative was evaluated using a linear spline interpolator.

    \item \textit{Single population with two-component broken power-law MF}. As before, only one isochrone is used to derive the LF, but the MF is allowed to have a break in the power law:

\begin{equation}
    \xi(M)\propto\begin{cases}
        M^{-\alpha_\mathrm{h}},& \text{if } M>M_\mathrm{pb}\\
        M^{-\alpha_\mathrm{l}},& \text{if } M\leq M_\mathrm{pb}\\
    \end{cases}
\label{eq:broken_MF}
\end{equation}

    where $\alpha_\mathrm{h}$ and $\alpha_\mathrm{l}$ are the power law slopes above and below the break, respectively, and $M_\mathrm{pb}$ is the stellar mass at which the break occurs.

    \item \textit{Mixture of two populations with identical single-component power-law MFs}. This family of models utilizes the single-component MF in Eq.\,\ref{eq:unbroken_MF}; however, the LF is composed of two populations with MLRs adopted from two distinct isochrones in panel (b) of Fig.\,\ref{fig:chem}. The contributions of the two populations to $\phi(m)$ are added with weights given by $\gamma$ and $(1-\gamma)$, where $0\leq\gamma\leq1$ is the mixing ratio. Both populations are assumed to have identical MFs.

    \item \textit{Mixture of two populations with identical two-component broken power-law MFs}. As before, but the shared MF of the two populations has a power law break (Eq.\,\ref{eq:broken_MF}).

    \item \textit{Mixture of two populations with distinct single-component power-law MFs}. These models consider two populations with single-component MFs in Eq.\,\ref{eq:unbroken_MF}; however, the slope of the power law is allowed to vary between the populations. We denote the slope of the MF of population 1 as $\alpha_\mathrm{h}$, and the slope of population 2 as $\alpha_2$.

\end{enumerate}

We did not consider more complicated LF models (e.g., two-population models with distinct broken power-law MFs or three-population models) to avoid over-fitting the observations; however, three-population configurations and arbitrary functional forms of the MF are explored in Section\,\ref{Section6.2}, where the observed CMD sequences are analysed separately.

The logarithmic likelihood in Eq.\,~\ref{eq:likelihood} was maximized using the Goodman-Weare \citep{MCMC} Markov Chain Monte Carlo (MCMC) sampling with respect to the free parameters defined above ($\alpha_\mathrm{h}$, $\alpha_\mathrm{l}$, $\alpha_2$, $M_\mathrm{pb}$ and $\gamma$, as appropriate for each family of LF models). We used $32$ walkers and $3000$ accepted steps per walker. The first $10\%$ of the accepted steps were discarded as burn-in. For single-population LF models, we ran separate MCMC chains for each of the four isochrones in panel (b) of Fig.\,\ref{fig:chem}. For LF models with two populations, we considered every possible pair of isochrones, except the ones where both populations have $\mathrm{Y}=0.4$, as such configurations are clearly unphysical. Furthermore, we ran separate chains for the LFs in F150W2 and F322W2 bands to verify the internal consistency of our results. The best-fit values of the free parameters were taken as the medians of the corresponding MCMC posteriors. The upper/lower $1$-sigma asymmetric errors were taken as half-differences between the median and the $97.7$/$2.3$ percentiles. The results are summarised in Table\,\ref{tab:MF_params}.

\begin{centering} 
\begin{figure}[!h]
 \includegraphics[width=\columnwidth]{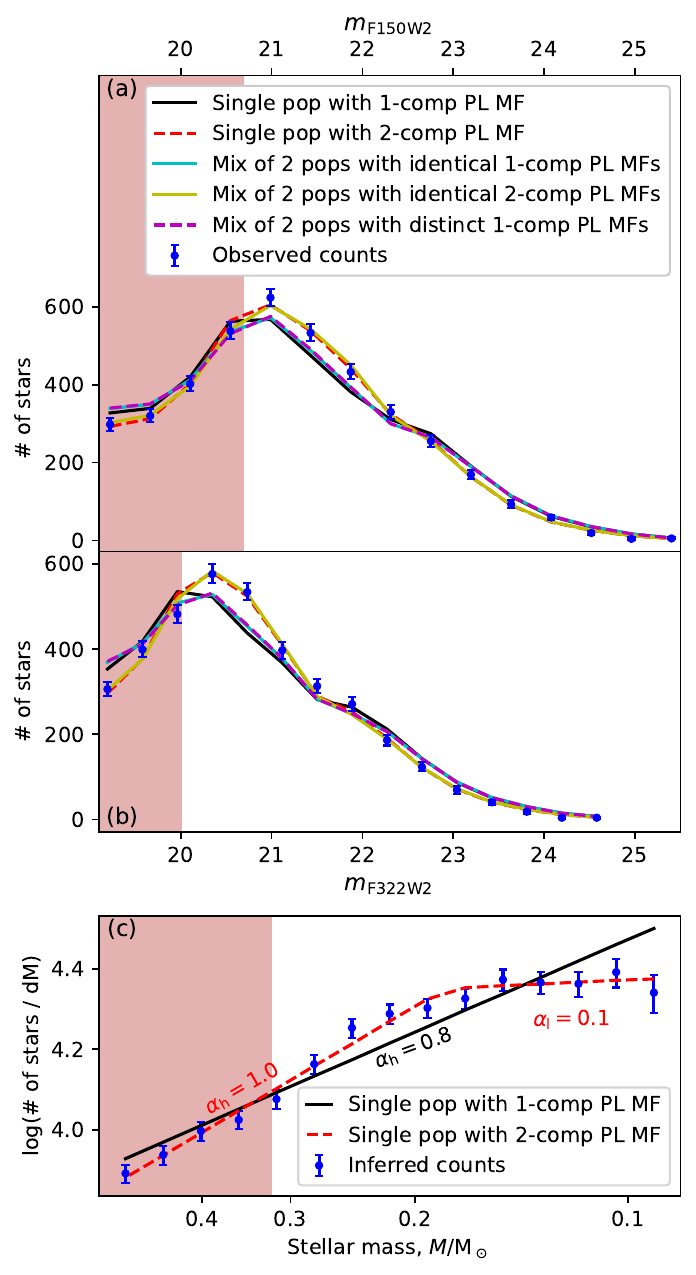}
 \caption{Panels (a) and (b) show the observed LFs of $\omega$\,Cen in F150W2 and F322W2 bands, respectively. The LFs are represented as histograms with $15$ uniform bins. Over-plotted in both panels are the best-fitting LF models (taken to be the model with the smallest GoF in F150W2) from each of the five families listed in Table\,\ref{tab:MF_params}. The model curves include completeness effects. Panel (c) shows the MF of the cluster based on the inferred stellar masses from the observed F150W2 magnitudes and the $\mathrm{Y}=0.25$ NOM isochrone, shown in panel (b) of Fig.\,\ref{fig:chem}. Unlike panels (a) and (b), the inferred counts in panel (c) have been corrected for photometric completeness. Over-plotted are the best-fitting single-component and two-component power-law MFs. In the legends, \textit{pop} stands for \textit{population}, \textit{comp} for \textit{component}, and \textit{PL} for \textit{power law}. The light-red areas qualitatively mark magnitude ranges where MS stars are mildly saturated in at least one filter. The LF and MF values in these regions should be interpreted with caution.}
 \label{LF_MF} 
\end{figure} 
\end{centering}

We note that our model fitting method does not require binning of data, and is free of issues associated with histogram-based approaches (e.g., \citealt{omega_cen_IMF,2024AN....34540039B,SANDee}) such as the dependence of results on the bin size and incorrect bin placements due to photometric errors. However, it is possible to visualise the best-fit model in the histogram form by integrating the completeness-corrected LF ($\phi(m_i)\ \mathrm{comp}(m_i)$ in Eq.\,\ref{eq:likelihood}) in the chosen magnitude bins. This has been carried out in $15$ uniform magnitude bins to produce panels (a) and (b) of Figure\,\ref{LF_MF}. These panels show the observed member counts in each bin, as well as the best-fitting (based on the maximum likelihood in the MCMC chain for F150W2) LF models from each of the five families described above. In all cases, the best-fitting models are the ones that use the $\mathrm{Y}=0.25$ NOM isochrone (for single-population models) or both $\mathrm{Y}=0.25$ and $\mathrm{Y}=0.4$ NOM isochrones (for mixed population models). The errors shown in Figure\,\ref{LF_MF} were estimated from the scatter among a large number of synthetic datasets that were generated using the best-fit LF model and the estimated completeness and photometric errors of our observations.

We also derived the goodness-of-fit (GoF) for each model as the average absolute difference between the observed and modelled histogram counts, normalised by the estimated count errors. The GoF of each considered LF model is shown in Table\,\ref{tab:MF_params}. Note that for $15$ magnitude bins, assuming Gaussian random errors and no systematic errors, the expected GoF is $\approx0.80\pm0.16$\footnote{Estimated as $\int \left|x\right|Z(x) dx\pm\sqrt{\int x^2 Z(x) dx - \left[\int \left|x\right|Z(x) dx\right]^2}/\sqrt{15}$, where $Z(x)$ is the standard normal distribution}. Models with GoFs significantly above/below this value are likely under/over-fitted.

The key conclusions that can be drawn from Table\,\ref{tab:MF_params} and Fig.\,\ref{LF_MF} are summarized below:

\begin{enumerate}
    \item Out of single-population models with unbroken single-component MFs, NOM25 and NOM40 provide the best fits to the data (smallest GoF) in F150W2 and F322W2, respectively. The latter case is clearly unphysical, since the majority of members in the observed field are expected to belong to rMS that has a near-solar $\mathrm{Y}\approx 0.25$. As we argue later in this section, this contradiction most likely arises because $\omega$\,Cen does have a break in the MF within the observed mass range ($\sim0.1-0.5\ \mathrm{M}_\odot$), and because super-solar values of $\mathrm{Y}$ mimic such a break in the LF (higher helium mass fractions create a discontinuity in the MLR near $0.3\ \mathrm{M}_\odot$, where stars become fully convective). If we only consider single-population unbroken-MF models with $\mathrm{Y}=0.25$, then NOM25 provides the best fit in both F150W2 and F322W2. The inferred MF slopes in both bands ($\sim 0.7-0.8$) are consistent within uncertainties and are also consistent with similar analyses in the literature (e.g., \citealt{omega_cen_IMF,2022ApJ...930...24G}).
    \item LF models with broken power-law MFs provide better fits to the data than the unbroken-MF models for all considered configurations. In most cases, adding a break to the MF improves the GoF by a factor of $\sim2$, and brings it in line with the expected value of $\approx0.80\pm0.16$ for a well-fitting model. Out of single-population broken-MF models, NOM25 offers the best fit in both F150W2 and F322W2. In both bands, a power law break around $M_\mathrm{pb}\approx0.2\ \mathrm{M}_\odot$ is observed with a steep bottom-heavy MF above the break ($\alpha_\mathrm{h}>1$ at $M>M_\mathrm{pb}$) and a nearly flat MF below the break ($\alpha_\mathrm{l}<0.2$ at $M\leq M_\mathrm{pb}$). Panel (c) of Fig.\,\ref{LF_MF} shows a histogram of the completeness-corrected inferred stellar masses using the NOM25 broken-MF model. The break in the power law is clearly seen.
    \item As emphasised above, the apparent break in the MF may actually be the result of a discontinuity in the MLR due to the transition from partially to fully convective stellar interiors. In our models, this transition occurs near $0.32\ \mathrm{M}_\odot$ for $\mathrm{Y}=0.25$, and $0.27\ \mathrm{M}_\odot$ for $\mathrm{Y}=0.4$. We however note that the best-fit $M_\mathrm{pb}$ for the broken-MF NOM25 model was estimated as $0.19\pm0.03\ \mathrm{M}_\odot$, i.e. over three sigma below the transition. This further corroborates that the observed MF break is likely genuine. On the other hand, all broken-MF single-population models with $\mathrm{Y}=0.4$ place the break near $0.3\ \mathrm{M}_\odot$. For these models, the inferred break is an artefact caused by the discontinuity in the MLR due to the transition from partially to fully convective stellar interiors.
    \item While introducing a break in the MF improves the GoF considerably, adding a second population to the mixture has a very small effect. This is apparent in panels (a) and (b) of Fig.\,\ref{LF_MF}, where the broken LF models with single and mixed populations are nearly indistinguishable from each other. 
    \item The best-fitting mixed population models are the combinations of NOM25+NOM40 and NOM25+HMHA25 populations with a common two-component broken MF. In the former case, the mixing ratio was estimated as $0.7\pm0.1$, suggesting that approximately $20-40\%$ of cluster members are expected to be significantly helium-enriched (bMS). This fraction is consistent with the fraction of observed bMS at this distance from the centre of $\omega$\,Cen \citep{2024A&A...688A.180S,2009A&A...507.1393B}. This result is a significant improvement over a similar analysis of \textit{HST} observations in \citet{2022ApJ...930...24G}, where only an upper limit on the population mixing ratio could be derived. In the following sub-section, we will verify the calculated mixing ratio by analysing the individual sequences of $\omega$\,Cen separately.
    \item The best-fitting broken-MF models in both single-population and mixed configurations reach the target GoF for a well-fitting model ($\approx0.80\pm0.16$). Addition of further degrees of freedom to the model would result in over-fitting. For this reason, we do not consider mixed models with distinct two-component MFs, or mixed models with more than $2$ populations. A far larger sample size is required to explore these more sophisticated models.
    \item Allowing a two-population LF model to have distinct MFs in each population has a very small effect on the GoF, compared to the corresponding LF models with identical single-component MFs. For nearly all distinct-MF models, the mixing ratio was found to be consistent with $\gamma=1$ within $2$-sigma bounds, and no reliable constraints on $\alpha_2$ could be derived. A notable exception is the NOM25+HMHA40 mixture, for which all parameters are well-constrained, and the best-fit values are consistent between F150W2 and F322W2. In this model, both populations were found to have similar MF slopes ($\alpha_\mathrm{h}\approx\alpha_2\approx 0.5$), which are comparable to the MF slopes calculated for single-population models with $\mathrm{Y}=0.4$. It is therefore likely that the best-fit parameters of the NOM25+HMHA40 model are an artefact of the discontinuity in the MLR at the transition between fully and partially convective stellar interiors.
\end{enumerate}

\subsection{Luminosity and mass functions for individual stellar populations}
\label{Section6.2}

In Section\,\ref{Section6.1}, we analysed the combined LF and MF of $\omega$\,Cen (where "combined" refers to the local LF and MF obtained by merging all subpopulations). However, this analysis did not fully exploit all the information available in the CMD, particularly the colour of each star relative to the overall colour distribution, which provides an indication of its likely population membership. As a result, we were limited in our ability to explore more complex models (such as two-population scenarios with distinct broken power-law MFs, or three-population models) and to examine in detail the differences in MF slopes among the individual populations. To overcome these limitations, we adopted a more direct approach, focusing on the LF and MF of the individual stellar populations over a narrower mass range, where the sequences are partially distinguishable. Unlike the method used in Section\,\ref{Section6.1}, here we adopted a binned approach, as it allows for a more accurate separation of the individual sequences and facilitates a clearer comparison of their respective LFs.

\begin{centering} 
\begin{figure*}
\centering
 \includegraphics[width=0.9\textwidth]{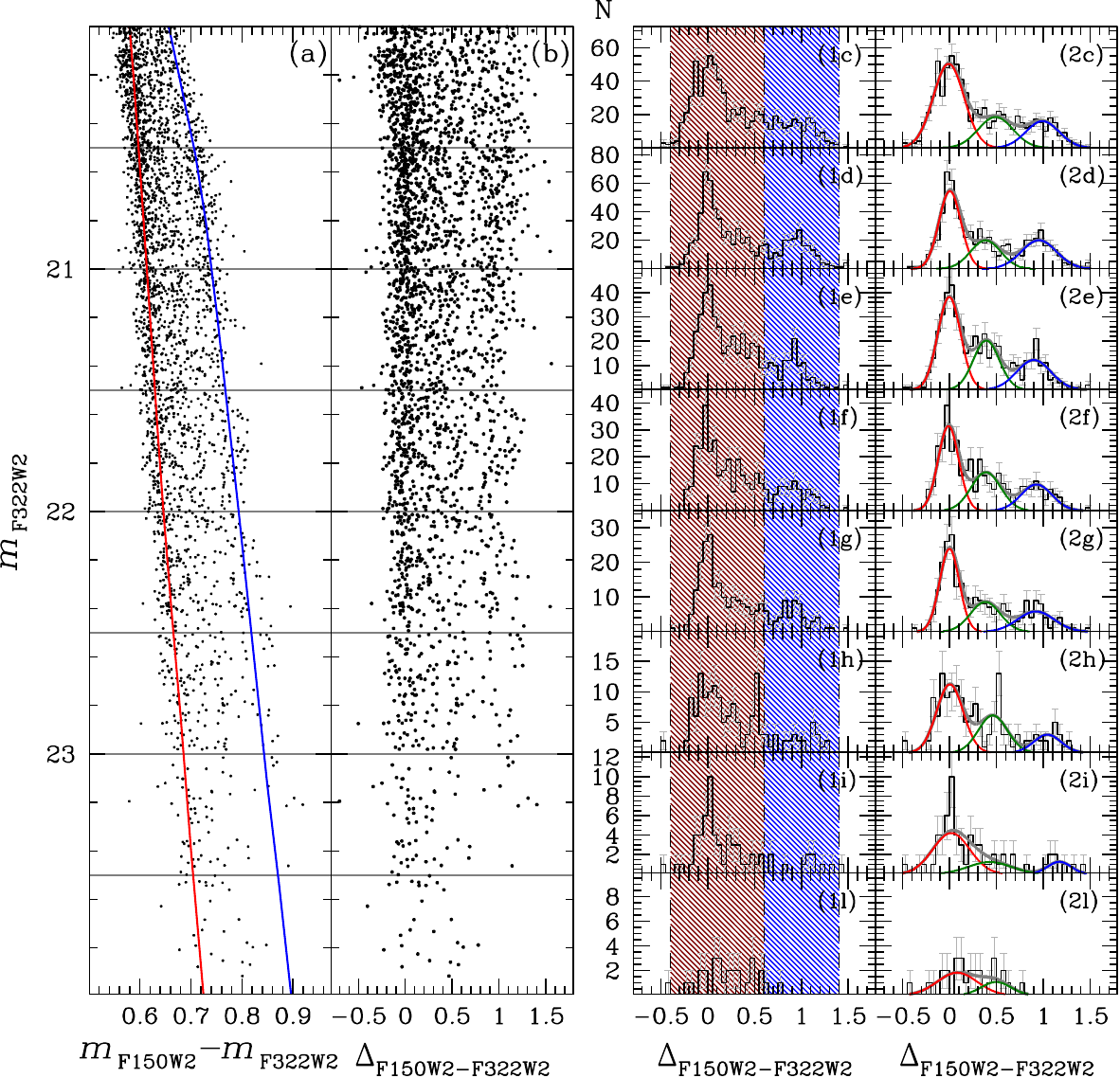}
 \caption{Procedure for estimating the LFs of the sequences in $\omega$\,Cen in the low-MS. Panel\,(a) shows the $m_{\rm F322W2}$ versus $m_{\rm F150W2}-m_{\rm F322W2}$ CMD, focusing on the region where the separation between the components is most evident. The red and blue lines indicate the fiducials of the rMS and bMS, respectively, and are used to construct the verticalised CMD shown in panel\,(b). Panels\,(1,2c) through (1,2l) display the histogram distribution of $\Delta_{m_{\rm F150W2}-m_{\rm F322W2}}$ for stars across eight magnitude intervals, as defined by the grey lines in panels\,(a) and (b). In panels\,(1c) to (1l), the red and blue regions corresponding to the two sequences are highlighted. These regions are used to estimate the LFs presented in Fig.\,\ref{LF1}, with values reported in Table\,\ref{tab:LF1}. Panels (2c) to (2l) present the same histograms with the best-fitting multi-Gaussian models overlaid: three-Gaussian fits in panels (2c) to (2i) and a two-Gaussian fit in panel (2l). The individual Gaussian components are shown in red, green, and blue, while the combined fit is plotted in grey. The obtained LFs are shown in Fig.\,\ref{LF2} and the values are reported in Table\,\ref{tab:LF2}.} 
 \label{3G} 
\end{figure*} 
\end{centering}

\begin{centering} 
\begin{figure}
 \includegraphics[width=\columnwidth]{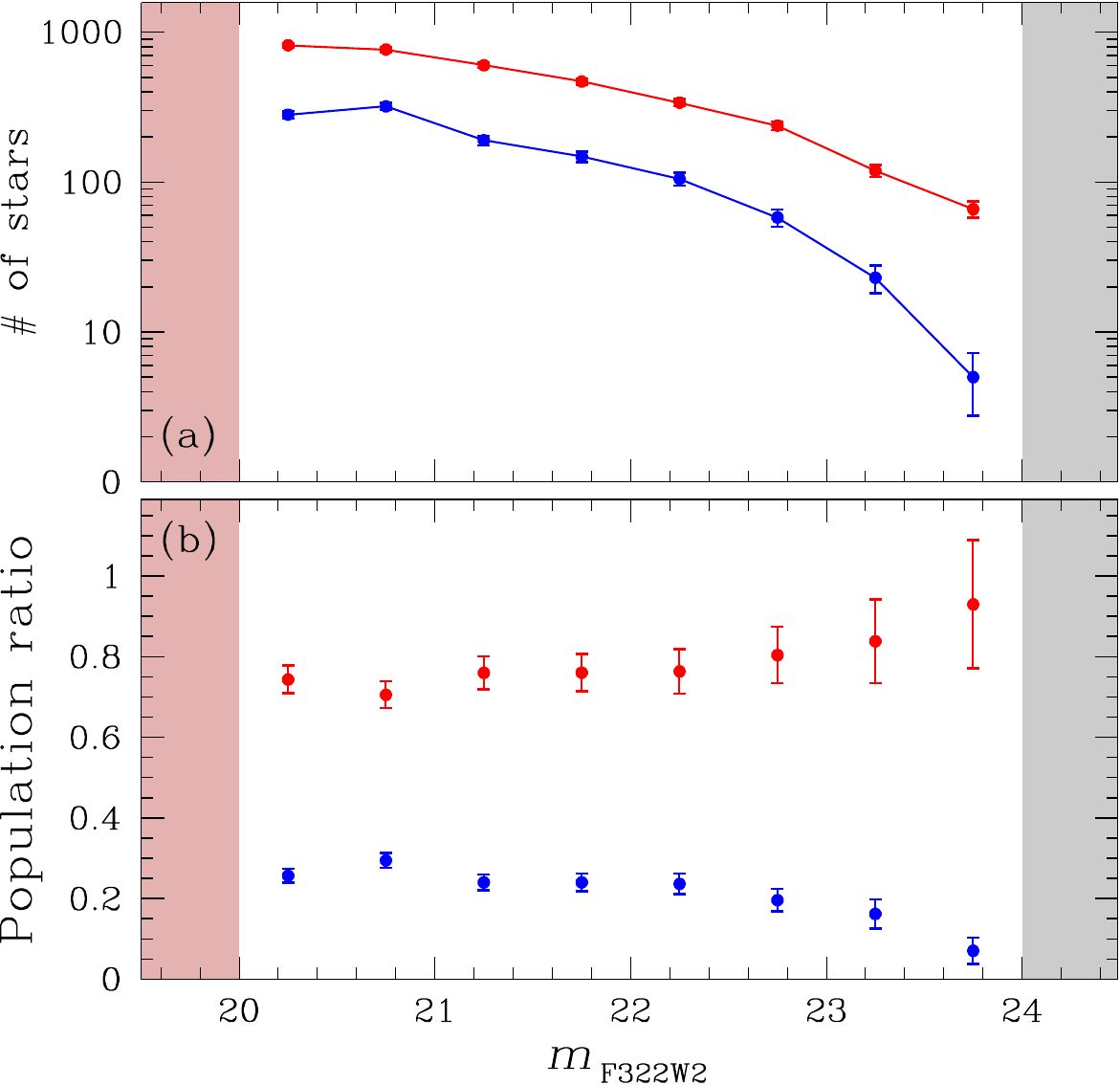}
 \caption{(a) LF derived using the regions defined in panels\,(1c) to (1l) of Fig.\,\ref{3G}. (b) Population ratio of stars associated with the two sequences. In both panels, the red-shaded region indicates the saturation limit, while the grey-shaded region marks the threshold beyond which the three populations can no longer be distinguished.} 
 \label{LF1} 
\end{figure} 
\end{centering}

We derived the LFs along the low-MS of the two populations, applying a methodology similar to that outlined in \citet{2024AN....34540018S}. The method is illustrated in Fig.\,\ref{3G}. Panel\,(a) of Fig.\,\ref{3G} shows the $m_{\rm F322W2}$ versus $m_{\rm F150W2}-m_{\rm F322W2}$ CMD for our selected sample of stars, in the magnitude range $20 < m_{\rm F322W2} < 24$, where the separation of the two components is most evident. Note that although completeness remains reliable down to $m_{\rm F322W2} \sim 26$, the statistical limitation of the sample prevents us from classifying or distinguishing additional sequences beyond  $m_{\rm F322W2} \sim 24$ (see Fig.\,\ref{CMDs}, \ref{CMDs2}, \ref{PMs}). We utilized the previously defined fiducials (see Fig\,\ref{crossing}) for the two sequences (shown in red and blue in panel\,(a) of Fig.\,\ref{3G}) to verticalise the diagram. For each star, we computed the verticalised colour as

\begin{equation}
\Delta_{\rm F150W2-F322W2} = \frac{X - X_{\rm red\,fiducial}}{X_{\rm blue\,fiducial} - X_{\rm red\,fiducial}}    
\end{equation}

where $X = m_{\rm F150W2} - m_{\rm F322W2}$ and $X_{\rm red\,fiducial}$, $X_{\rm blue\,fiducial}$ are are the colours of the red and blue fiducials, respectively, evaluated at the star’s $m_{\rm F322W2}$ magnitude. The resulting verticalised diagram is shown in panel\,(b) of Fig.\,\ref{3G}. We defined 8 magnitude bins in $m_{\rm F322W2}$, each 0.5 magnitudes wide (indicated by the grey horizontal lines in panels\,(a) and (b)). The histogram of the verticalised colour for each magnitude bin is shown in panels\,(1,2c) to (1,2l). We defined two regions: one with $-0.4<\Delta_{m_{\rm F150W2}-m_{\rm F322W2}}<0.6$ corresponding to the rMS, and another with $0.6<\Delta_{m_{\rm F150W2}-m_{\rm F322W2}}<1.4$ corresponding to the bMS. These regions are represented in panels (1c) to (1l) in red and blue, respectively. For each bin, we counted the number of stars within each of the two defined regions. The resulting values, corrected for completeness, are shown in panel\,(a) of Fig.\,\ref{LF1}, using the same colour scheme as in Fig.\,\ref{3G}, with error bars representing Poisson errors (this convention for the error bars is adopted throughout this figure and in all subsequent LF and MF plots). As observed, the two LFs exhibit a similar shape, with the rMS containing a larger number of stars in each magnitude interval. The bMS LF declines more rapidly than the rMS LF at fainter magnitudes ($m_{\rm F322W2} > 22.5$) due to the earlier termination of the bMS sequence. 

Panel\,(b) of Fig.\,\ref{LF1} presents the population ratio between the two sequences. The error bars represent uncertainties derived through standard error propagation -- this approach is also adopted in all subsequent figures displaying population ratios. We calculated the weighted mean of the ratios, finding that the rMS accounts for $75\%\pm1\%$, while the bMS accounts for $25\%\pm2\%$ of the total number of stars. These values are consistent with those previously reported in literature \citep[][bMS/rMS$\sim$0.33, for a radial distance corresponding to the field analysed in this study, $\sim2-3$r$_h$]{2024A&A...688A.180S,2009A&A...507.1393B}, and agree with the results discussed in Section\,\ref{Section6.1}. The data presented in Fig.\,\ref{LF1} are listed in Table\,\ref{tab:LF1}.

To estimate the LFs of the two sequences while accounting for possible contamination between them, we used an alternative method, detailed in Section\,4.2 of \citet{2024AN....34540018S}. This method involves determining the number of stars in each bin and in each sequence by fitting the histograms with a multi-Gaussian model, where the number of Gaussians corresponds to the number of sequences. In our case, we initially aimed to fit the histograms with two Gaussian components, one for each sequence. However, upon closer inspection of the verticalised diagram in panel\,(b) and the histograms in panels\,(1c, 2c) to (1l, 2l), we observed evidence of a possible third component situated between the two main sequences. As discussed in the introduction, $\omega$\,Cen hosts a highly complex system of mPOPs. While the bMS and rMS constitute the dominant populations, additional stellar groups have been identified, each exhibiting substructures (including the rMS and bMS), resulting in at least 15 distinct populations within $\omega$\,Cen \citep{2017ApJ...844..164B,2024A&A...688A.180S}. The MS presented in panels\,(a) and (b), and the relative histograms shown in panels\,(1,2c) to (1,2l), are the result of the contribution of all these stellar populations, which overlap and blend with each other. This complexity necessitates a more precise method to determine the optimal number of Gaussian components to fit.

To address this, we applied a Gaussian mixture model (GMM) to the $\Delta_{m_{\rm F150W2}-m_{\rm F322W2}}$ distribution of stars in each bin, using the expectation-maximisation algorithm from the scikit-learn package \citep{2011JMLR...12.2825P}. To find the best model, we calculated the Bayesian Information Criterion (BIC) for models with one to five Gaussian components. A three-Gaussian model provided the lowest BIC for most bins, except for the last two bins (those with, respectively, 23$<m_{\rm F322W2}<$23.5 and 23.5$<m_{\rm F322W2}<$24) where a two-Gaussian and a one-Gaussian model were optimal, respectively. In these final bins, the rightmost sequence is barely detectable in the penultimate bin and entirely absent in the last, leaving only the leftmost sequence in the last bin, along with a faint remnant of the middle sequence. Based on these results, we adopted a three-Gaussian model for all bins except the last one, where a two-Gaussian model was used. The GMM fits are shown in grey in panels\,(2c) to (2l), with the individual components represented in red, green, and blue. Hereafter, we will refer to the leftmost and rightmost components as the rMS and bMS, respectively, while the middle sequence will be referred to as the green MS (gMS).

\begin{centering} 
\begin{figure}
 \includegraphics[width=\columnwidth]{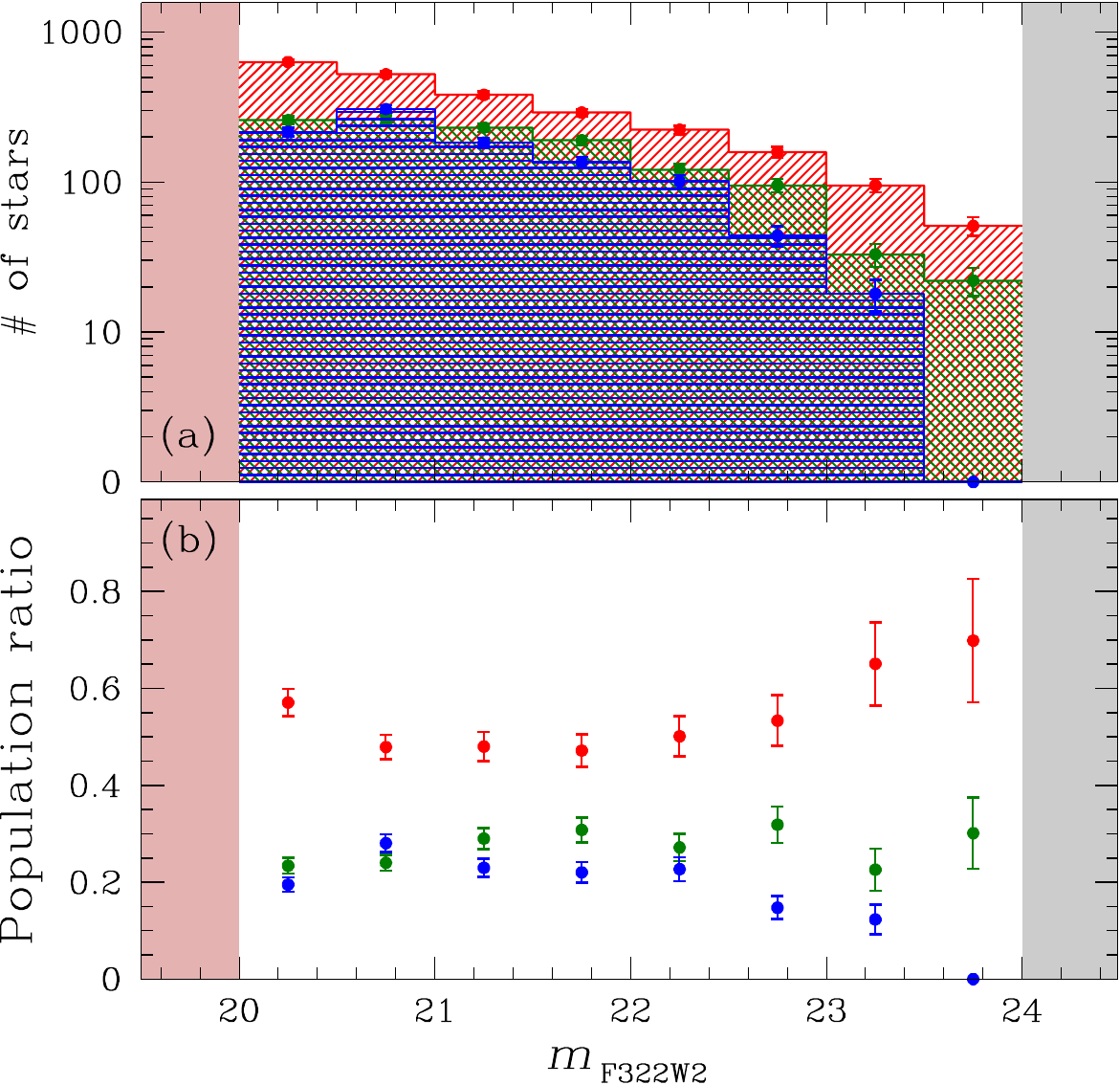}
 \caption{(a) LF obtained from the three-Gaussian fit shown in panels (2c) to (2l) of Fig.\,\ref{3G}, highlighting the drop of the bMS in the final bin ($m_{\rm F150W2} \sim 23.75$). (b) Population ratio of stars associated with the three sequences. As in Fig.\,\ref{LF1}, both panels feature a red-shaded region indicating the saturation limit and a grey-shaded region marking the threshold beyond which the three populations can no longer be distinguished.} 
 \label{LF2} 
\end{figure} 
\end{centering}

For each bin, the number of stars in each sequence was estimated from the area under the corresponding Gaussian component. The resulting LFs, corrected for completeness, are illustrated in panel\,(a) of Fig.\,\ref{LF2}. As shown, the three LFs follow a similar shape. The rMS is the most populous sequence, the bMS is the least populous, and the gMS lies in between. The only exception occurs at $m_{\rm F322W2}=20.75$, where a slight peak in the bMS LF temporarily exceeds that of gMS. In the lower part, the LF for the bMS shows a drop, reaching zero in the final bin as the sequence seems to end. Panel\,(b) shows the population ratios of the three sequences across each magnitude bin. We calculated the weighted mean of the ratios, finding that the rMS accounts for $52\%\pm2\%$, the gMS for $26\%\pm1\%$, and the bMS for $22\%\pm2\%$ of the total number of stars. The data presented in Fig.\,\ref{LF2} are listed in Table\,\ref{tab:LF2}.

Table\,\ref{tab:LF2} also reports the dispersions ($\sigma$) of the three Gaussian components used to model the colour distribution of the MS stars. To assess whether the observed colour spread of the three components arises from intrinsic star-to-star variations in C and O abundances -- rather than being solely due to photometric errors -- we performed a similar analysis on the ASs shown in panel\,(a) of Fig.\,\ref{ASs}. We verticalised the two ASs sequences (black points in panel\,(a) of Fig.\,\ref{ASs}) using the same fiducials as for the real stars, and divided the verticalised CMD into 0.5-mag bins in $m_{\rm F322W2}$, over the same magnitude range used for the real data ($20 < m_{\rm F322W2} < 24$). In each bin, we applied the GMM to estimate the colour dispersion of the two components. The resulting $\sigma$ values are listed in Table\,\ref{tab:LF2}. These dispersions are systematically smaller than those measured from the real data, except in the two faintest bins, where photometric errors start to become significant -- particularly for the bMS. It is worth noting that the dispersions for the two injected sequences are very similar across the full magnitude range, diverging only in the last two bins. This comparison confirms that the colour dispersion observed in the real stars cannot be explained by photometric errors alone, and is consistent with intrinsic variations in chemical composition, most notably in C and O abundances.

As pointed out by the Referee, the red tail in the artificial stars discussed in Section\,\ref{Section5} could, in principle, contribute to the apparent presence of the gMS. However, we note that such red tails involve only a small fraction of the recovered stars, which is insufficient to account for the number of stars observed in the gMS. Moreover, both the colour dispersion and the extent of these tails are significantly smaller than those measured in the real data, particularly in the region occupied by the gMS. We therefore find no indication that the gMS could be the result of this effect.

\begin{table}[h!]
\caption{Number of stars in the two sequences (rMS and bMS) and their relative ratios, obtained using the method illustrated in panels\,(1c) to (1l) of Fig.\,\ref{3G}. The final row reports the total number of stars and the weighted mean of the ratios for the two sequences.}
\centering
\begin{tabular}{c|cc|cc}
\hline
\textbf{$m_{\rm F322W2}$} & \textbf{rMS} & \textbf{bMS} & \textbf{rMS/N} & \textbf{bMS/N} \\
 & & & \% & \% \\
\hline
$20.0 - 20.5$ & 817$\pm$29 & 282$\pm$17 & 74$\pm$3 & 26$\pm$2 \\
$20.5 - 21.0$ & 766$\pm$28 & 320$\pm$18 & 71$\pm$3 & 29$\pm$2 \\
$21.0 - 21.5$ & 602$\pm$25 & 190$\pm$14 & 76$\pm$4 & 24$\pm$2 \\
$21.5 - 22.0$ & 469$\pm$22 & 148$\pm$12 & 76$\pm$5 & 24$\pm$2 \\
$22.0 - 22.5$ & 339$\pm$18 & 105$\pm$10 & 76$\pm$6 & 24$\pm$3 \\
$22.5 - 23.0$ & 238$\pm$15 & 58$\pm$8  & 80$\pm$7  & 20$\pm$3 \\
$23.0 - 23.5$ & 119$\pm$11 & 23$\pm$5  & 84$\pm$10 & 16$\pm$4 \\
$23.5 - 24.0$ & 66$\pm$8   & 5$\pm$2   & 93$\pm$16 & 7$\pm$3 \\
\hline
 & 3416$\pm$58  & 1132$\pm$34 & 75$\pm$1 & 25$\pm$2 \\
\hline
\end{tabular}
\label{tab:LF1}
\end{table}

\begin{table*}[h!]
\caption{Values derived from the three-Gaussian fit method illustrated in panels\,(1c) to (1l) of Fig.\,\ref{3G} for the three sequences. The table also includes the dispersion values ($\sigma$) for each Gaussian component for both the real and artificial stars. The final row presents the total number of stars and the weighted average of the ratios for the three sequences.}
\centering
\begin{tabular}{c|ccc|ccc|ccc|cc}
\hline
\textbf{$m_{\rm F322W2}$} & \textbf{rMS} & \textbf{gMS} & \textbf{bMS} & \textbf{rMS/N} & \textbf{gMS/N} & \textbf{bMS/N}  & \textbf{$\sigma_{\rm rMS}$} & \textbf{$\sigma_{\rm gMS}$} & \textbf{$\sigma_{\rm bMS}$}  & \textbf{$\sigma_{\rm rMS}^{\rm ART}$} & \textbf{$\sigma_{\rm bMS}^{\rm ART}$}\\
 & & & & \% & \% & \% & & & & & \\
\hline
$20.0 - 20.5$ & 632$\pm$25 & 259$\pm$16 & 216$\pm$15 & 57$\pm$3  & 23$\pm$2 & 20$\pm$1 & 0.16 & 0.18 & 0.18 & 0.02 & 0.02\\
$20.5 - 21.0$ & 524$\pm$23 & 263$\pm$16 & 307$\pm$18 & 48$\pm$3  & 24$\pm$2 & 28$\pm$2 & 0.12 & 0.16 & 0.19 & 0.02 & 0.02\\
$21.0 - 21.5$ & 382$\pm$20 & 231$\pm$15 & 183$\pm$14 & 48$\pm$3  & 29$\pm$2 & 23$\pm$2 & 0.12 & 0.14 & 0.18 & 0.03 & 0.03\\
$21.5 - 22.0$ & 291$\pm$17 & 190$\pm$14 & 136$\pm$12 & 47$\pm$3  & 31$\pm$3 & 22$\pm$2 & 0.11 & 0.16 & 0.16 & 0.03 & 0.04\\
$22.0 - 22.5$ & 223$\pm$15 & 121$\pm$11 & 101$\pm$10 & 50$\pm$4  & 27$\pm$3 & 23$\pm$3 & 0.10 & 0.16 & 0.19 & 0.05 & 0.05\\
$22.5 - 23.0$ & 159$\pm$13 &  95$\pm$10 &  44$\pm$7  & 53$\pm$5  & 32$\pm$4 & 15$\pm$2 & 0.14 & 0.15 & 0.15 & 0.07 & 0.07\\
$23.0 - 23.5$ &  95$\pm$10 &  33$\pm$6  &  18$\pm$4  & 65$\pm$9  & 23$\pm$4 & 12$\pm$3 & 0.20 & 0.23 & 0.13 & 0.08 & 0.11\\
$23.5 - 24.0$ &  51$\pm$7  &  22$\pm$5  &   0$\pm$0  & 70$\pm$13 & 30$\pm$7 &  0$\pm$0 & 0.21 & 0.16 & 0.00 & 0.08 & 0.14\\
\hline
 & 2357$\pm$49 & 1214$\pm$35 & 1005$\pm$32 & 52$\pm$2 & 26$\pm$1 & 22$\pm$2 & & & \\
\hline
\end{tabular}
\label{tab:LF2}
\end{table*}

To derive the MFs of the three stellar sequences, we adopted a statistical approach. In each magnitude bin defined in Fig.\,\ref{3G}, we randomly selected three sub-samples of stars and assigned them to the three populations. The number of stars in each sub-sample was determined according to the area under the corresponding Gaussian component. The magnitudes of the stars were then converted into masses using the isochrones shown in Fig.\,\ref{fig:chem}. In particular, we adopted the NOM isochrone with $\mathrm{Y}=0.25$ for stars assigned to the rMS and gMS, and the NOM isochrone with $\mathrm{Y}=0.40$ for those assigned to the bMS (see Tables\,\ref{tab:NOM25} and \ref{tab:NOM40} for the corresponding tabulated MLRs). We then computed the MF for each sequence by counting the number of stars in logarithmic mass bins. This entire procedure was repeated 1000 times, and the final MFs were obtained by taking the median number of stars in each mass bin across all realisations.

\begin{centering} 
\begin{figure}
 \includegraphics[width=\columnwidth]{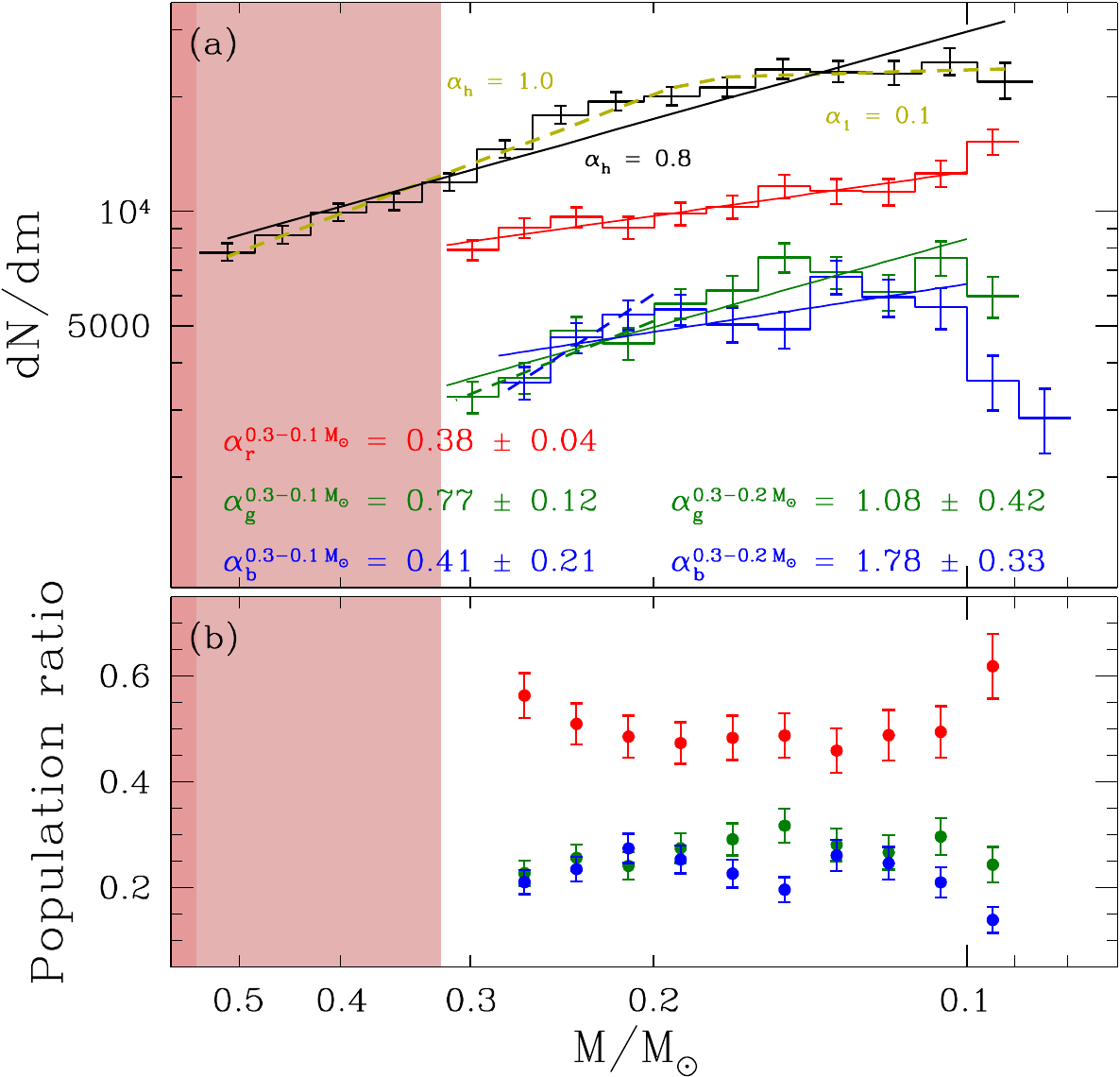}
 \caption{a) MFs of the three identified sequences, along with interpolating lines in the mass range $0.1 < \mathrm{M} < 0.3$\,M$_{\odot}$ (solid lines). The corresponding power-law slopes are indicated in the plot. For the gMS and bMS, we also show fits in the narrower mass range $0.2 < \mathrm{M} < 0.3$\,M$_{\odot}$ (dashed lines), with their respective slopes also reported. The black histogram represents the combined MF, as shown in panel\,(c) of Fig.\,\ref{LF_MF}. The solid black line and dashed yellow line correspond to the single-component and broken (two-component) power-law fits to the combined MF, respectively, with their slopes provided for reference. (b) Population ratio of stars associated with the three sequences as a function of stellar mass. The light-red and dark-red shaded regions highlight areas affected by mild and severe saturation, respectively.}
 \label{MF} 
\end{figure} 
\end{centering}

The resulting MFs for each sequence are shown in Fig.\,\ref{MF}, alongside the combined MF previously presented in panel\,(c) of Fig.\,\ref{LF_MF}. The MFs of the individual populations cover a mass range of $\sim$0.30–0.09\,M$_{\odot}$ for the rMS and gMS, and $\sim$0.27–0.08\,M$_{\odot}$ for the bMS. Each MF was fitted with a single power-law function (see Eq.\,\ref{eq:unbroken_MF}) over the mass range $0.1 < \mathrm{M} < 0.3$\,M$_{\odot}$. The best-fit lines and the corresponding slope values ($\alpha$) are shown in the figure.

Among the populations, the gMS exhibits the steepest slope ($\alpha_{\mathrm{g}} = 0.77 \pm 0.12$), while the rMS and bMS show similar and flatter slopes ($\alpha_{\mathrm{r}} = 0.38 \pm 0.04$ and $\alpha_{\mathrm{b}} = 0.41 \pm 0.14$, respectively). The MF of the rMS -– the most populated sequence -– is well characterized by a single power-law, indicating a reliable MLR. In contrast, the MFs of the gMS and bMS are noisier and display a noticeable flattening around $\sim$0.2\,M$_{\odot}$, which also causes the flattening observed in the combined MF at the same mass. This behaviour could indicate a real drop in the number of stars formed below $\sim$0.2\,M$_{\odot}$, or it might reflect uncertainties in the adopted MLR, particularly at low masses. It is also interesting to notice that for masses smaller than 0.2\,M$_{\odot}$, both the gMS and the bMS are characterized a MF steeper than that of the rMS. This steep slope is responsible for the steepening of the combined MF.

It is important to note that dividing the MS into only three components represents a simplification, given the high level of complexity in the stellar populations of $\omega$\,Cen. Moreover, the sequence identified as the rMS in the optical HST CMD evolves into a broader colour distribution in the NIR JWST CMD at magnitudes fainter than $m_{\rm F150W2} \simeq 20$ (or $m_{\rm F322W2} \simeq 19$; see Fig.\,\ref{crossing}), with a main peak corresponding to the red Gaussian component (Fig.\,\ref{3G}) and an extended tail that overlaps with the other two components. These features suggest a higher level of complexity in the stellar population structure, and a more refined analysis would require a more sophisticated and detailed approach.

Finally, panel\,(b) displays the population ratios of the three sequences across each mass bin. We computed the weighted mean of these ratios, finding that the rMS comprises $50\%\pm1\%$ of the total number of stars, while the gMS and bMS account for $27\%\pm1\%$ and $23\%\pm1\%$, respectively. These values are consistent with those derived from the LFs (see Fig.\,\ref{LF2}) and listed in Table\,\ref{tab:LF2}. The data presented in Fig.\,\ref{MF} are listed in Table\,\ref{tab:MF}.

\begin{table*}[h!]
\caption{Mass functions values for the three sequences and populations ratios, as shown in Fig.\,\ref{MF}. The final row presents the weighted average of the ratios for the three sequences.}
\centering
\begin{tabular}{c|ccc|ccc}
\hline
\textbf{M/M$_{\odot}$} & \textbf{rMS} & \textbf{gMS} & \textbf{bMS} & \textbf{rMS/N} & \textbf{gMS/N} & \textbf{bMS/N} \\
 & & & & \% & \% & \% \\
\hline
$0.316 - 0.282$ &  7,909$\pm$480   & 3,254$\pm$308 & -             & -        & -        & -        \\
$0.282 - 0.251$ &  9,060$\pm$544   & 3,651$\pm$345 & 3,546$\pm$348 & 56$\pm$4 & 23$\pm$2 & 21$\pm$2 \\
$0.251 - 0.224$ &  9,661$\pm$595   & 4,858$\pm$422 & 4,661$\pm$423 & 51$\pm$4 & 26$\pm$2 & 23$\pm$2 \\
$0.224 - 0.199$ &  9,057$\pm$610   & 4,494$\pm$430 & 5,355$\pm$480 & 49$\pm$4 & 24$\pm$3 & 27$\pm$3 \\
$0.199 - 0.178$ &  9,867$\pm$674   & 5,724$\pm$514 & 5,521$\pm$516 & 47$\pm$4 & 28$\pm$3 & 25$\pm$3 \\
$0.178 - 0.158$ & 10,285$\pm$729   & 6,200$\pm$566 & 5,047$\pm$523 & 48$\pm$4 & 29$\pm$3 & 23$\pm$3 \\
$0.158 - 0.141$ & 11,645$\pm$822   & 7,572$\pm$663 & 4,903$\pm$546 & 49$\pm$4 & 32$\pm$3 & 19$\pm$2 \\
$0.141 - 0.126$ & 11,308$\pm$858   & 6,920$\pm$671 & 6,728$\pm$677 & 46$\pm$4 & 28$\pm$3 & 26$\pm$3 \\
$0.126 - 0.112$ & 11,263$\pm$907   & 6,143$\pm$670 & 5,943$\pm$674 & 49$\pm$5 & 27$\pm$3 & 24$\pm$3 \\
$0.112 - 0.100$ & 12,596$\pm$1,016 & 7,548$\pm$786 & 5,595$\pm$693 & 49$\pm$5 & 30$\pm$4 & 21$\pm$3 \\
$0.100 - 0.089$ & 15,237$\pm$1,184 & 5,986$\pm$742 & 3,582$\pm$587 & 62$\pm$6 & 24$\pm$3 & 14$\pm$2 \\
$0.089 - 0.076$ & -                & -             & 2,863$\pm$556 & -        & -        & -        \\
\hline
 & & & & 50$\pm$1 & 27$\pm$1 & 23$\pm$1 \\
\hline
\end{tabular}
\label{tab:MF}
\end{table*}

\subsubsection{Note on the effect of the helium mass fraction}

As shown in Section\,\ref{Section4}, the stars of the bMS population are likely helium-enriched compared to their rMS counterparts with the difference in $\mathrm{Y}$ between the two populations reaching $\sim0.15$. In general, a star with enriched helium content has a higher mean molecular weight in the interior, and is more luminous than an equally massive star with a near-solar $\mathrm{Y}$. However, a helium-enriched population of stars would also be able to sustain nuclear fusion at lower stellar masses, thereby lowering the hydrogen-burning limit. These two effects largely suppress each other, making the lower MS almost completely insensitive to $\mathrm{Y}$.

In other words, the observed LF is determined not by the MLR, but by the slope of the MLR with respect to stellar mass (see Eq.\,\ref{eq:MF_to_LF}). While $\mathrm{Y}$ has a large impact on the MLR, its effect on the slope is subtle compared to the effect of e.g. atmospheric chemistry. For this reason, it is unlikely that the high helium content of the bMS is responsible for its early termination in the CMD compared to the rMS. This is further illustrated in Table\,\ref{tab:HBL}, where we provide the hydrogen-burning limits and their corresponding magnitudes for the isochrones considered in this study. Note that the large effect of $\mathrm{Y}$ on the former does not propagate to the latter.

\begin{table}[h!]
\caption{Hydrogen-burning limits (HBLs) and corresponding magnitudes of model isochrones.}
\centering
\begin{tabular}{c|ccc}
\hline
  & & &  \\
Isochrone & HBL[${M_\odot}$] & $m_{\rm F150W2}$ & $m_{\rm F322W2}$ \\
  & & &  \\
\hline
  & & &  \\
\texttt{NOM25} & 0.084 & 27.84 & 26.19 \\
\texttt{NOM40} & 0.070 & 27.81 & 26.27 \\
\texttt{HMHA25} & 0.080 & 28.11 & 26.37 \\
\texttt{HMHA40} & 0.066 & 28.09 & 26.49 \\
  & & &  \\
\hline
\end{tabular}
\label{tab:HBL}
\end{table}

\section{Summary}\label{Section7}

We presented the first study of the most massive GC in our Galaxy, $\omega$\,Cen, using recently acquired JWST proprietary data \citep[GO-5110][]{2024jwst.prop.5110B}. The results of our investigation are summarised as follows.

We constructed the cluster's NIR CMDs using JWST photometry. Figure\,\ref{final} displays these CMDs across broad colour and magnitude ranges, with the onset of mild and strong saturation clearly marked (see Section\,\ref{Section2} for further details). In these CMDs, two sequences, -- corresponding to the optical bMS and rMS populations of the cluster -- can be clearly separated from $m_{\rm F150W2} \sim18$ ($m_{\rm F322W2} \sim17.5$) down to $m_{\rm F150W2} \sim 20$ ($m_{\rm F322W2} \sim 19.5$), where they intersect and change their relative positions (see Fig.\,\ref{crossing}). Below this intersection, the bMS shifts toward redder colours, while the rMS extends over a broad colour range, predominantly favouring bluer colours. The sequences remain distinguishable down to the end of the MS. The sequence corresponding to the bMS stars appears to end -- at least -- $\sim$0.5\,magnitudes brighter than the sequence associated with the rMS population. We performed AS tests to evaluate the completeness of our data and assess whether the observed sequence terminations are intrinsic or driven by incompleteness. The results clearly indicate that the earlier termination of the bMS, which occurs at significantly brighter magnitudes compared to the rMS, is an intrinsic property of the population and not an artefact of completeness effects.

Our comparison with theoretical isochrones confirms that the colour spread in the CMD is primarily driven by variations in helium ($\mathrm{Y}$) abundance above the MS knee and by atmospheric chemistry below the knee. We found that a scatter in $\mathrm{Y}$ of approximately 0.15 is necessary to fully capture the observed width of the CMD above the MS knee, reinforcing the idea that helium enhancement plays a dominant role in shaping the stellar populations of $\omega$\,Cen.

Below the MS knee, the broader colour distribution is largely influenced by variations in oxygen and carbon abundances, which regulate the strength of $\mathrm{H_2O}$ absorption bands in the infrared. While the observed spread can be approximated by models with a range of $[\mathrm{C/Fe}]$ and $[\mathrm{O/Fe}]$, reproducing the full extent of the red tail requires extreme chemical compositions that are unlikely given the known abundance correlations in globular clusters. Instead, we find that a combination of metallicity variations and light element abundance changes provides a more plausible explanation for the observed CMD morphology.

By combining our JWST data with archival HST observations, as recently presented in \citet{2024A&A...691A..96S}, we calculated PMs and conducted a membership analysis. We then derived the combined LF and MF of $\omega$\,Cen, independent of individual populations, by fitting a forward model to the observed magnitudes of confirmed members using an MCMC method. We considered five different families of LF models: (1) a single population with a single-component power-law MF, (2) a single population with a two-component broken power-law MF, (3) a mixture of two populations with identical single-component power-law MFs, (4) a mixture of two populations with identical two-component broken power-law MFs, and (5) a mixture of two populations with distinct single-component power-law MFs. We explored various combinations of isochrones for the populations and evaluated their contributions to the LF.

Our results indicate that the single-population broken power-law models, particularly those using the $\mathrm{Y}=0.25$ NOM isochrone, provide the best fit. The MF exhibits a break around $0.2\ \mathrm{M}_\odot$, with a steep slope above the break and a flatter slope below it. Adding a second population did not substantially improve the fit. The observed LF is predominantly influenced by a single population with an average chemical composition. The mixed population models suggest that approximately 20-40\% of cluster members may be helium-enriched.

We then analysed the LF and MF of individual MS populations in $\omega$\,Cen by focusing on a narrower mass range ($\sim$0.3-0.08\,M$_{\odot}$) where the sequences are more clearly separated. The LFs of both sequences exhibit similar shapes, with the rMS containing a larger fraction of stars, accounting for $75\%\pm1\%$ of the total population. The bMS LF declines more rapidly than the rMS LF at fainter magnitudes ($m_{\rm F322W2} > 22.5$) due to the earlier termination of the bMS sequence. The bMS-to-rMS number ratio of $\sim0.33$ in this external field agrees with values previously reported by \citet{2024A&A...688A.180S,2009A&A...507.1393B} for fields at a similar radial distance from the cluster centre.
    
To refine our LF estimates and account for potential contamination between the sequences, we adopted an alternative approach. We analysed the verticalised CMD by fitting the star distributions in different magnitude bins with a multi-Gaussian model. By closely inspecting the verticalised CMD, we identified evidence of a possible third sequence between the bMS and rMS. To determine the optimal number of Gaussian components for each magnitude bin, we applied a GMM. The GMM results favoured three Gaussian components for most bins. The LFs were then estimated based on the area under each Gaussian component.

The derived LFs show similar shapes for all three components. In the lower part of the LFs, the bMS exhibits a drop, reaching zero in the final bin. The rMS is the most populous, contributing $52\%\pm2\%$ of the stars, followed by the gMS at $26\%\pm1\%$, and the bMS at $22\%\pm2\%$.

We then estimated the MF for each of the three sequences in the mass range between 0.3\,M$_{\odot}$ and 0.08\,M$_{\odot}$. The rMS is well-fitted by a single power-law in this mass range. The gMS and the bMS, on the other hand, are characterised by a flattening in their MFs for masses <0.2\,M$_{\odot}$ and a slope for masses >0.2\,M$_{\odot}$ that, for both populations, is steeper than that of the rMS. The steepening of the combined MF for masses smaller larger than 0.2\,M$_{\odot}$ and the flattening at masses <0.2\,M$_{\odot}$ are linked with the variation of the MF of the gMS and bMS. The flattening around $\sim$0.2\,M$_{\odot}$ for the gMS and the bMS may reflect a real drop in the number of low-mass stars or might be due to uncertainties in the adopted MLR. The gMS and bMS MF steeper than the rMS for masses >0.2\,M$_{\odot}$ may be the result of the effects of internal two-body relaxation on the radial variation of the MF for the two populations initially more centrally concentrated than the rMS \citep[see][for further discussion and details; see also e.g. \citealt{2024MNRAS.534.2397L} for the effects of two-body relaxation on the evolution towards energy equipartition and Ziliotto et al.\,(2025, submitted), for an observational study of energy equipartition in $\omega$\,Cen]{2018MNRAS.476.2731V}. We emphasize however that the narrow mass range and uncertainties in the MF derivation call for caution in the interpretation of the observational results. Additional studies covering a broader mass range will be necessary to further investigate the differences between the MF of multiple populations in $\omega$\,Cen.

This work represents the first study of the multiple populations in $\omega$\,Cen using JWST data. As is well known, $\omega$\,Cen hosts up to 15 distinct stellar populations in its core. In this study, we focused on the LFs of the sequences identifiable with JWST photometry. However, the complexity of the multiple populations in this cluster requires a more in-depth analysis to fully disentangle and characterise its stellar populations. This is particularly crucial when combining JWST and HST photometry, as demonstrated in a recent study \citep{2024A&A...689A..59S}. A comprehensive investigation utilising this combined approach will be the subject of a forthcoming paper (Gerasimov et al.\, 2025, in preparation).

\begin{acknowledgements}
Michele Scalco and Luigi Rolly Bedin acknowledge support by MIUR under the PRIN-2017 programme \#2017Z2HSMF,  by INAF under the PRIN-2019 programme \#10-Bedin, and by INAF under the WFAP project, f.o.:1.05.23.05.05. JA, AB, and EV acknowledge support from JWST program GO-5110. We thank our Referee, Dr. Ricardo Salinas, for his prompt and insightful review of our manuscript. His constructive comments and suggestions have significantly improved the clarity and quality of this work. This research is based on observations made with the NASA/ESA Hubble Space Telescope and the NASA/ESA/CSA James Webb Space Telescope. The data were obtained from the Mikulski Archive for Space Telescopes (MAST) at the Space Telescope Science Institute (STScI), which is operated by the Association of Universities for Research in Astronomy, Inc., under NASA contracts NAS 5–26555 for HST and NAS 5-03127 for JWST. These observations are associated with programs HST-GO-14118+14662 and JWST-GO-5110. The data described here may be obtained from \url{http://dx.doi.org/10.17909/7prx-4905}.
\end{acknowledgements}

\bibliographystyle{aa}
\bibliography{main.bib}

\begin{appendix}
\section{Overview of the JWST photometric catalogue}\label{Appendix:A}

\begin{centering} 
\begin{figure}
 \includegraphics[width=\columnwidth]{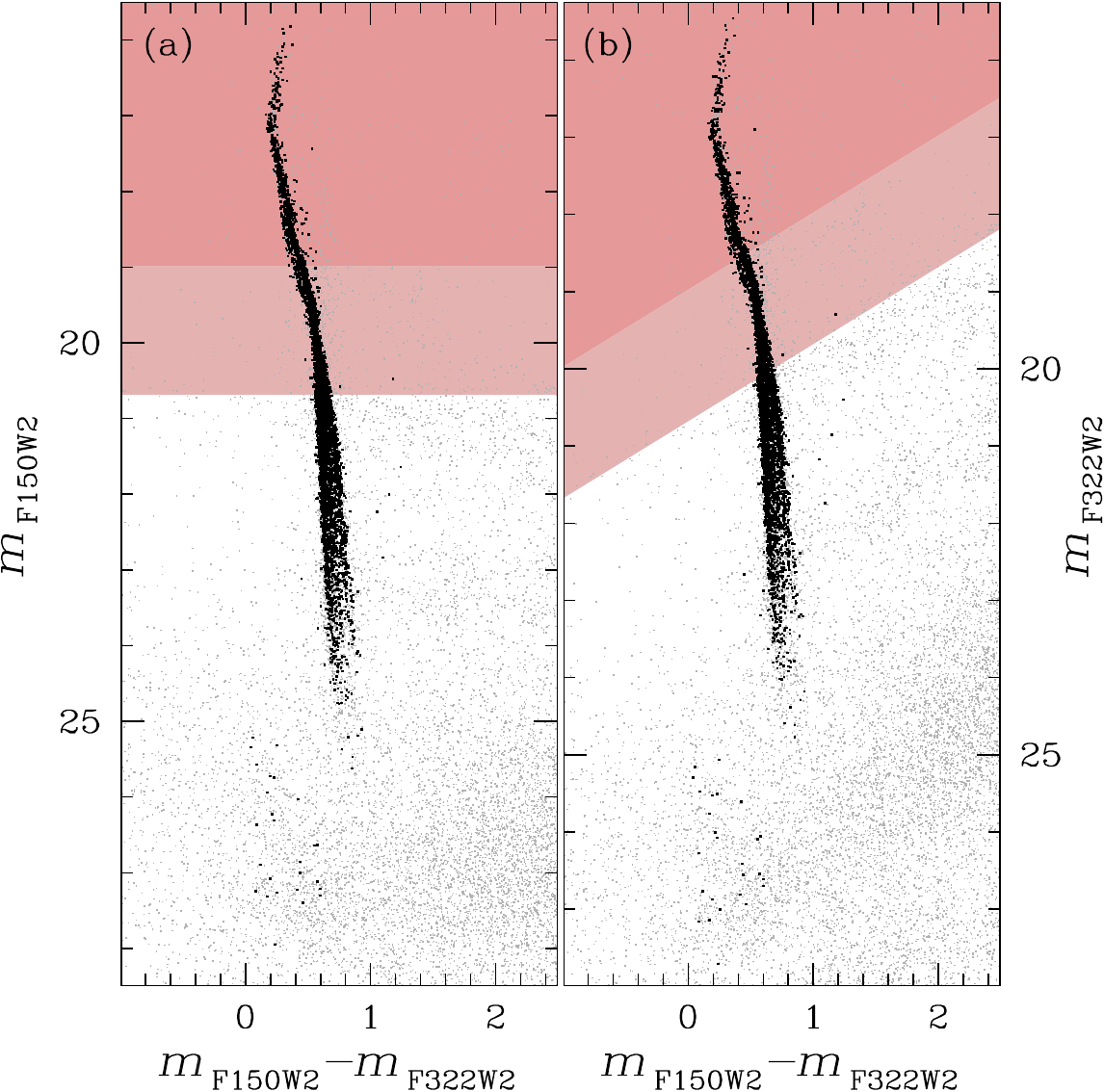}
 \caption{Colour-magnitude diagrams of $\omega$\,Cen using JWST filters. Panel (a) shows the $m_{\rm F150W2}$ versus $m_{\rm F150W2}-m_{\rm F322W2}$ CMD, while panel (b) displays the $m_{\rm F322W2}$ versus $m_{\rm F150W2}-m_{\rm F322W2}$ CMD, including all stars within the JWST FOV. In both panels, black dots represent stars that pass both the photometric quality and PM selection criteria, while grey dots indicate the remaining stars. The light-red and dark-red shaded regions highlight areas affected by saturation: the light-red region corresponds to saturated photometry, while the dark-red region marks severe saturation, where even \texttt{frame\,zero} is saturated.} 
 \label{final} 
\end{figure} 
\end{centering}

\section{Mass-luminosity relations}\label{Appendix:B}

\begin{table*}
    \centering
        \caption{Mass-luminosity relation from the NOM25 model. The table lists stellar masses, effective temperatures, and magnitudes in the F150W2 and F322W2 filters, both with and without reddening corrections.}
    \begin{tabular}{cccccccccccc}
        \cmidrule(lr){1-6} \cmidrule(lr){7-12}
        M/M$_{\odot}$ & T$_{\rm eff}$ [K] & $m_{\rm F150W2}$ & $m_{\rm F322W2}$ & $m_{\rm F150W2}^{\rm reddened}$ & $m_{\rm F322W2}^{\rm reddened}$ & M/M$_{\odot}$ & T$_{\rm eff}$ [K] & $m_{\rm F150W2}$ & $m_{\rm F322W2}$ & $m_{\rm F150W2}^{\rm reddened}$ & $m_{\rm F322W2}^{\rm reddened}$\\
        \cmidrule(lr){1-6} \cmidrule(lr){7-12}
    0.0835 & 1326 & 28.5716 & 26.6109 & 28.6539 & 26.6285 & 0.1277 & 3618 & 22.6685 & 22.0451 & 22.7410 & 22.0663 \\
    0.0836 & 1343 & 28.5085 & 26.5738 & 28.5909 & 26.5916 & 0.1372 & 3674 & 22.4687 & 21.8517 & 22.5412 & 21.8729 \\
    0.0839 & 1406 & 28.2747 & 26.4414 & 28.3575 & 26.4592 & 0.1484 & 3739 & 22.2525 & 21.6417 & 22.3249 & 21.6630 \\
    0.0839 & 1428 & 28.1914 & 26.3977 & 28.2745 & 26.4156 & 0.1604 & 3813 & 22.0333 & 21.4287 & 22.1058 & 21.4501 \\
    0.0842 & 1503 & 27.9088 & 26.2544 & 27.9930 & 26.2724 & 0.1729 & 3874 & 21.8317 & 21.2317 & 21.9042 & 21.2531 \\
    0.0844 & 1579 & 27.6186 & 26.0950 & 27.7035 & 26.1132 & 0.1871 & 3904 & 21.6542 & 21.0575 & 21.7267 & 21.0789 \\
    0.0844 & 1609 & 27.5086 & 26.0312 & 27.5937 & 26.0494 & 0.2062 & 3933 & 21.4502 & 20.8591 & 21.5227 & 20.8806 \\
    0.0846 & 1693 & 27.2394 & 25.8559 & 27.3244 & 25.8742 & 0.2284 & 3961 & 21.2432 & 20.6572 & 21.3157 & 20.6787 \\
    0.0847 & 1780 & 26.9732 & 25.6822 & 27.0580 & 25.7007 & 0.2539 & 3990 & 21.0370 & 20.4557 & 21.1095 & 20.4772 \\
    0.0849 & 1864 & 26.7293 & 25.5259 & 26.8141 & 25.5446 & 0.2831 & 4017 & 20.8320 & 20.2569 & 20.9046 & 20.2784 \\
    0.0850 & 1944 & 26.5138 & 25.3814 & 26.5983 & 25.4002 & 0.3170 & 4047 & 20.6235 & 20.0574 & 20.6961 & 20.0790 \\
    0.0852 & 2036 & 26.2801 & 25.2189 & 26.3642 & 25.2378 & 0.3559 & 4089 & 20.3942 & 19.8395 & 20.4668 & 19.8611 \\
    0.0854 & 2108 & 26.0932 & 25.0886 & 26.1770 & 25.1076 & 0.3968 & 4162 & 20.1166 & 19.5763 & 20.1893 & 19.5979 \\
    0.0857 & 2184 & 25.9133 & 24.9337 & 25.9960 & 24.9529 & 0.4309 & 4258 & 19.8574 & 19.3369 & 19.9303 & 19.3585 \\
    0.0861 & 2288 & 25.6578 & 24.7009 & 25.7390 & 24.7205 & 0.4604 & 4376 & 19.6168 & 19.1223 & 19.6900 & 19.1440 \\
    0.0864 & 2357 & 25.5094 & 24.5908 & 25.5903 & 24.6104 & 0.4797 & 4474 & 19.4566 & 18.9837 & 19.5300 & 19.0053 \\
    0.0870 & 2451 & 25.3086 & 24.4470 & 25.3890 & 24.4666 & 0.4853 & 4506 & 19.4092 & 18.9430 & 19.4827 & 18.9646 \\
    0.0875 & 2541 & 25.1170 & 24.2954 & 25.1967 & 24.3151 & 0.5036 & 4619 & 19.2596 & 18.8168 & 19.3334 & 18.8384 \\
    0.0881 & 2622 & 24.9556 & 24.1539 & 25.0344 & 24.1738 & 0.5200 & 4734 & 19.1294 & 18.7090 & 19.2036 & 18.7307 \\
    0.0889 & 2712 & 24.7795 & 23.9945 & 24.8574 & 24.0145 & 0.5350 & 4845 & 19.0128 & 18.6131 & 19.0873 & 18.6348 \\
    0.0897 & 2798 & 24.6034 & 23.8373 & 24.6804 & 23.8575 & 0.5491 & 4955 & 18.9047 & 18.5244 & 18.9796 & 18.5462 \\
    0.0907 & 2882 & 24.4436 & 23.6943 & 24.5199 & 23.7146 & 0.5545 & 4998 & 18.8643 & 18.4914 & 18.9393 & 18.5131 \\
    0.0919 & 2968 & 24.2908 & 23.5560 & 24.3664 & 23.5764 & 0.5677 & 5107 & 18.7650 & 18.4103 & 18.8403 & 18.4320 \\
    0.0930 & 3036 & 24.1605 & 23.4381 & 24.2356 & 23.4586 & 0.5804 & 5213 & 18.6702 & 18.3326 & 18.7459 & 18.3543 \\
    0.0946 & 3116 & 23.9920 & 23.2858 & 24.0665 & 23.3064 & 0.5928 & 5318 & 18.5765 & 18.2553 & 18.6524 & 18.2770 \\
    0.0966 & 3192 & 23.8194 & 23.1285 & 23.8934 & 23.1492 & 0.6052 & 5422 & 18.4819 & 18.1764 & 18.5581 & 18.1981 \\
    0.0993 & 3268 & 23.6395 & 22.9628 & 23.7130 & 22.9837 & 0.6176 & 5522 & 18.3850 & 18.0942 & 18.4614 & 18.1159 \\
    0.1028 & 3344 & 23.4515 & 22.7879 & 23.5247 & 22.8088 & 0.6299 & 5621 & 18.2853 & 18.0086 & 18.3619 & 18.0304 \\
    0.1073 & 3418 & 23.2567 & 22.6049 & 23.3296 & 22.6259 & 0.6425 & 5720 & 18.1813 & 17.9185 & 18.2582 & 17.9402 \\
    0.1129 & 3490 & 23.0610 & 22.4192 & 23.1338 & 22.4403 & 0.6551 & 5818 & 18.0719 & 17.8223 & 18.1490 & 17.8441 \\
    0.1196 & 3556 & 22.8667 & 22.2345 & 22.9393 & 22.2557 & 0.6680 & 5915 & 17.9547 & 17.7177 & 18.0319 & 17.7395 \\
        \cmidrule(lr){1-6} \cmidrule(lr){7-12}
    \end{tabular}
    \label{tab:NOM25}
\end{table*}

\begin{table*}
    \centering
        \caption{Same as \ref{tab:NOM25} but for the NOM40 model.}
    \begin{tabular}{cccccccccccc}
        \cmidrule(lr){1-6} \cmidrule(lr){7-12}
        M/M$_{\odot}$ & T$_{\rm eff}$ [K] & $m_{\rm F150W2}$ & $m_{\rm F322W2}$ & $m_{\rm F150W2}^{\rm reddened}$ & $m_{\rm F322W2}^{\rm reddened}$ & M/M$_{\odot}$ & T$_{\rm eff}$ [K] & $m_{\rm F150W2}$ & $m_{\rm F322W2}$ & $m_{\rm F150W2}^{\rm reddened}$ & $m_{\rm F322W2}^{\rm reddened}$\\
        \cmidrule(lr){1-6} \cmidrule(lr){7-12}
    0.0700 & 1363 & 28.6613 & 26.7627 & 28.7439 & 26.7805 & 0.1172 & 3704 & 22.6225 & 22.0087 & 22.6950 & 22.0300 \\
    0.0701 & 1413 & 28.4807 & 26.6603 & 28.5637 & 26.6782 & 0.1264 & 3770 & 22.4039 & 21.7959 & 22.4764 & 21.8173 \\
    0.0702 & 1485 & 28.2132 & 26.5222 & 28.2971 & 26.5401 & 0.1361 & 3875 & 22.1658 & 21.5654 & 22.2383 & 21.5868 \\
    0.0703 & 1506 & 28.1338 & 26.4814 & 28.2180 & 26.4994 & 0.1452 & 3908 & 22.0041 & 21.4076 & 22.0766 & 21.4290 \\
    0.0704 & 1593 & 27.7885 & 26.2917 & 27.8736 & 26.3098 & 0.1571 & 3935 & 21.8311 & 21.2401 & 21.9036 & 21.2615 \\
    0.0705 & 1688 & 27.4746 & 26.0877 & 27.5597 & 26.1061 & 0.1730 & 3965 & 21.6324 & 21.0471 & 21.7050 & 21.0686 \\
    0.0706 & 1780 & 27.1868 & 25.8976 & 27.2717 & 25.9161 & 0.1917 & 3996 & 21.4303 & 20.8503 & 21.5029 & 20.8718 \\
    0.0707 & 1866 & 26.9316 & 25.7314 & 27.0164 & 25.7501 & 0.2135 & 4025 & 21.2269 & 20.6543 & 21.2996 & 20.6758 \\
    0.0708 & 1949 & 26.7065 & 25.5774 & 26.7911 & 25.5962 & 0.2390 & 4058 & 21.0188 & 20.4552 & 21.0915 & 20.4767 \\
    0.0709 & 2034 & 26.4796 & 25.4172 & 26.5637 & 25.4361 & 0.2681 & 4099 & 20.8072 & 20.2548 & 20.8798 & 20.2764 \\
    0.0711 & 2104 & 26.2911 & 25.2853 & 26.3749 & 25.3043 & 0.2945 & 4148 & 20.5749 & 20.0319 & 20.6476 & 20.0535 \\
    0.0712 & 2168 & 26.1322 & 25.1473 & 26.2151 & 25.1664 & 0.3010 & 4165 & 20.5163 & 19.9765 & 20.5891 & 19.9981 \\
    0.0715 & 2252 & 25.9230 & 24.9583 & 26.0046 & 24.9777 & 0.3264 & 4253 & 20.2699 & 19.7481 & 20.3428 & 19.7696 \\
    0.0718 & 2315 & 25.7724 & 24.8269 & 25.8533 & 24.8464 & 0.3496 & 4370 & 20.0263 & 19.5300 & 20.0995 & 19.5517 \\
    0.0723 & 2400 & 25.5982 & 24.7058 & 25.6789 & 24.7254 & 0.3646 & 4468 & 19.8634 & 19.3887 & 19.9368 & 19.4103 \\
    0.0728 & 2495 & 25.3874 & 24.5513 & 25.4676 & 24.5710 & 0.3694 & 4502 & 19.8108 & 19.3435 & 19.8843 & 19.3651 \\
    0.0734 & 2575 & 25.2273 & 24.4148 & 25.3066 & 24.4346 & 0.3837 & 4617 & 19.6577 & 19.2141 & 19.7315 & 19.2358 \\
    0.0742 & 2668 & 25.0456 & 24.2522 & 25.1239 & 24.2721 & 0.3961 & 4727 & 19.5268 & 19.1048 & 19.6010 & 19.1265 \\
    0.0750 & 2757 & 24.8672 & 24.0921 & 24.9446 & 24.1122 & 0.4073 & 4833 & 19.4100 & 19.0078 & 19.4845 & 19.0295 \\
    0.0759 & 2843 & 24.6986 & 23.9416 & 24.7752 & 23.9618 & 0.4179 & 4937 & 19.3012 & 18.9175 & 19.3760 & 18.9393 \\
    0.0769 & 2931 & 24.5397 & 23.7988 & 24.6156 & 23.8192 & 0.4281 & 5040 & 19.1968 & 18.8307 & 19.2719 & 18.8524 \\
    0.0781 & 3018 & 24.3847 & 23.6586 & 24.4599 & 23.6791 & 0.4381 & 5141 & 19.0944 & 18.7450 & 19.1698 & 18.7667 \\
    0.0795 & 3097 & 24.2219 & 23.5120 & 24.2965 & 23.5326 & 0.4479 & 5243 & 18.9927 & 18.6596 & 19.0684 & 18.6813 \\
    0.0800 & 3122 & 24.1671 & 23.4622 & 24.2415 & 23.4829 & 0.4576 & 5344 & 18.8915 & 18.5741 & 18.9675 & 18.5958 \\
    0.0820 & 3207 & 23.9797 & 23.2917 & 24.0536 & 23.3125 & 0.4671 & 5444 & 18.7898 & 18.4873 & 18.8660 & 18.5090 \\
    0.0844 & 3284 & 23.8016 & 23.1280 & 23.8751 & 23.1488 & 0.4766 & 5543 & 18.6844 & 18.3965 & 18.7609 & 18.4183 \\
    0.0875 & 3361 & 23.6148 & 22.9542 & 23.6880 & 22.9752 & 0.4863 & 5642 & 18.5741 & 18.3004 & 18.6508 & 18.3221 \\
    0.0915 & 3439 & 23.4204 & 22.7714 & 23.4933 & 22.7925 & 0.4960 & 5740 & 18.4574 & 18.1973 & 18.5343 & 18.2191 \\
    0.0964 & 3513 & 23.2235 & 22.5851 & 23.2963 & 22.6062 & 0.5059 & 5839 & 18.3317 & 18.0849 & 18.4088 & 18.1067 \\
    0.1023 & 3582 & 23.0267 & 22.3984 & 23.0993 & 22.4196 & 0.5159 & 5937 & 18.1935 & 17.9593 & 18.2708 & 17.9811 \\
    0.1092 & 3646 & 22.8265 & 22.2064 & 22.8991 & 22.2276 & & & & & & \\
        \cmidrule(lr){1-6} \cmidrule(lr){7-12}
    \end{tabular}
    \label{tab:NOM40}
\end{table*}

\begin{table*}
    \centering
        \caption{Same as \ref{tab:NOM25} but for the HMHA25 model.}
    \begin{tabular}{cccccccccccc}
        \cmidrule(lr){1-6} \cmidrule(lr){7-12}
        M/M$_{\odot}$ & T$_{\rm eff}$ [K] & $m_{\rm F150W2}$ & $m_{\rm F322W2}$ & $m_{\rm F150W2}^{\rm reddened}$ & $m_{\rm F322W2}^{\rm reddened}$ & M/M$_{\odot}$ & T$_{\rm eff}$ [K] & $m_{\rm F150W2}$ & $m_{\rm F322W2}$ & $m_{\rm F150W2}^{\rm reddened}$ & $m_{\rm F322W2}^{\rm reddened}$\\
        \cmidrule(lr){1-6} \cmidrule(lr){7-12}
    0.0600 &  876 & 31.0094 & 27.9167 & 31.0943 & 27.9327 & 0.0951 & 2929 & 23.9720 & 23.3425 & 24.0459 & 23.3628  \\
    0.0655 &  939 & 30.3506 & 27.5939 & 30.4339 & 27.6101 & 0.0981 & 3011 & 23.7976 & 23.1743 & 23.8712 & 23.1948  \\
    0.0700 &  979 & 30.2239 & 27.3969 & 30.3002 & 27.4133 & 0.1017 & 3086 & 23.6214 & 23.0052 & 23.6948 & 23.0257  \\
    0.0725 & 1006 & 30.2140 & 27.3423 & 30.2874 & 27.3589 & 0.1064 & 3161 & 23.4335 & 22.8249 & 23.5068 & 22.8456  \\
    0.0739 & 1040 & 30.0181 & 27.2836 & 30.0934 & 27.3001 & 0.1120 & 3232 & 23.2440 & 22.6420 & 23.3171 & 22.6627  \\
    0.0760 & 1108 & 29.5327 & 27.0911 & 29.6105 & 27.1077 & 0.1188 & 3299 & 23.0544 & 22.4577 & 23.1274 & 22.4785  \\
    0.0774 & 1156 & 29.2349 & 26.9057 & 29.3113 & 26.9225 & 0.1268 & 3364 & 22.8582 & 22.2688 & 22.9312 & 22.2897  \\
    0.0779 & 1177 & 29.1027 & 26.8249 & 29.1785 & 26.8418 & 0.1361 & 3421 & 22.6637 & 22.0809 & 22.7366 & 22.1018  \\
    0.0787 & 1228 & 28.8396 & 26.6662 & 28.9148 & 26.6832 & 0.1472 & 3491 & 22.4478 & 21.8738 & 22.5206 & 21.8949  \\
    0.0790 & 1257 & 28.7180 & 26.5928 & 28.7930 & 26.6099 & 0.1588 & 3571 & 22.2286 & 21.6621 & 22.3015 & 21.6832  \\
    0.0796 & 1321 & 28.4066 & 26.4498 & 28.4833 & 26.4670 & 0.1710 & 3624 & 22.0395 & 21.4775 & 22.1123 & 21.4987  \\
    0.0797 & 1339 & 28.2870 & 26.4202 & 28.3655 & 26.4374 & 0.1858 & 3657 & 21.8524 & 21.2927 & 21.9252 & 21.3139  \\
    0.0802 & 1409 & 27.8620 & 26.3091 & 27.9465 & 26.3264 & 0.2046 & 3691 & 21.6417 & 21.0839 & 21.7144 & 21.1051  \\
    0.0805 & 1470 & 27.6884 & 26.1947 & 27.7727 & 26.2122 & 0.2261 & 3722 & 21.4322 & 20.8739 & 21.5049 & 20.8951  \\
    0.0806 & 1504 & 27.5988 & 26.1352 & 27.6829 & 26.1528 & 0.2506 & 3753 & 21.2224 & 20.6629 & 21.2951 & 20.6842  \\
    0.0809 & 1600 & 27.4362 & 25.9869 & 27.5188 & 26.0047 & 0.2786 & 3787 & 21.0128 & 20.4520 & 21.0854 & 20.4733  \\
    0.0811 & 1689 & 27.0805 & 25.8019 & 27.1641 & 25.8198 & 0.3102 & 3821 & 20.8053 & 20.2436 & 20.8778 & 20.2649  \\
    0.0814 & 1770 & 26.7972 & 25.6054 & 26.8800 & 25.6235 & 0.3463 & 3857 & 20.6110 & 20.0495 & 20.6835 & 20.0708  \\
    0.0816 & 1833 & 26.6014 & 25.4628 & 26.6834 & 25.4809 & 0.3791 & 3908 & 20.3703 & 19.8092 & 20.4427 & 19.8306  \\
    0.0819 & 1911 & 26.3859 & 25.3085 & 26.4673 & 25.3268 & 0.3890 & 3925 & 20.3020 & 19.7417 & 20.3744 & 19.7632  \\
    0.0823 & 1982 & 26.2051 & 25.1914 & 26.2863 & 25.2098 & 0.4230 & 4002 & 20.0481 & 19.4936 & 20.1205 & 19.5150  \\
    0.0829 & 2093 & 25.9101 & 24.9861 & 25.9907 & 25.0046 & 0.4542 & 4069 & 19.7939 & 19.2798 & 19.8663 & 19.3013  \\
    0.0834 & 2181 & 25.6850 & 24.8170 & 25.7647 & 24.8358 & 0.4843 & 4179 & 19.5313 & 19.0432 & 19.6037 & 19.0648  \\
    0.0840 & 2264 & 25.4766 & 24.6566 & 25.5555 & 24.6755 & 0.5046 & 4277 & 19.3528 & 18.8782 & 19.4252 & 18.8998  \\
    0.0846 & 2347 & 25.2796 & 24.4991 & 25.3577 & 24.5181 & 0.5110 & 4310 & 19.2977 & 18.8282 & 19.3701 & 18.8498  \\
    0.0855 & 2432 & 25.0799 & 24.3394 & 25.1572 & 24.3587 & 0.5303 & 4426 & 19.1354 & 18.6855 & 19.2080 & 18.7072  \\
    0.0864 & 2516 & 24.8854 & 24.1833 & 24.9621 & 24.2027 & 0.5469 & 4535 & 19.0030 & 18.5718 & 19.0759 & 18.5936  \\
    0.0876 & 2600 & 24.6962 & 24.0137 & 24.7721 & 24.0334 & 0.5621 & 4641 & 18.8884 & 18.4754 & 18.9616 & 18.4971  \\
    0.0889 & 2683 & 24.5110 & 23.8463 & 24.5864 & 23.8662 & 0.5765 & 4746 & 18.7846 & 18.3894 & 18.8581 & 18.4111  \\
    0.0906 & 2765 & 24.3295 & 23.6824 & 24.4043 & 23.7024 & 0.5902 & 4849 & 18.6873 & 18.3094 & 18.7613 & 18.3311  \\
    0.0926 & 2847 & 24.1502 & 23.5144 & 24.2245 & 23.5346 & 0.6000 & 4923 & 18.6184 & 18.2526 & 18.6927 & 18.2743  \\
        \cmidrule(lr){1-6} \cmidrule(lr){7-12}
    \end{tabular}
    \label{tab:HMHA25}
\end{table*}

\begin{table*}
    \centering
        \caption{Same as \ref{tab:NOM25} but for the HMHA40 model.}
    \begin{tabular}{cccccccccccc}
        \cmidrule(lr){1-6} \cmidrule(lr){7-12}
        M/M$_{\odot}$ & T$_{\rm eff}$ [K] & $m_{\rm F150W2}$ & $m_{\rm F322W2}$ & $m_{\rm F150W2}^{\rm reddened}$ & $m_{\rm F322W2}^{\rm reddened}$ & M/M$_{\odot}$ & T$_{\rm eff}$ [K] & $m_{\rm F150W2}$ & $m_{\rm F322W2}$ & $m_{\rm F150W2}^{\rm reddened}$ & $m_{\rm F322W2}^{\rm reddened}$\\
        \cmidrule(lr){1-6} \cmidrule(lr){7-12}
    0.0600 &  991 & 30.4865 & 27.6025 & 30.5609 & 27.6189 & 0.0862 & 3091 & 23.8141 & 23.1989 & 23.8875 & 23.2194 \\
    0.0606 & 1000 & 30.5013 & 27.5873 & 30.5745 & 27.6038 & 0.0902 & 3170 & 23.6254 & 23.0183 & 23.6987 & 23.0390 \\
    0.0624 & 1053 & 30.1610 & 27.4707 & 30.2368 & 27.4873 & 0.0950 & 3244 & 23.4360 & 22.8356 & 23.5091 & 22.8564 \\
    0.0638 & 1114 & 29.7069 & 27.2730 & 29.7844 & 27.2897 & 0.1007 & 3316 & 23.2446 & 22.6504 & 23.3176 & 22.6713 \\
    0.0641 & 1132 & 29.5940 & 27.2027 & 29.6709 & 27.2195 & 0.1074 & 3379 & 23.0514 & 22.4642 & 23.1244 & 22.4852 \\
    0.0649 & 1177 & 29.3136 & 27.0312 & 29.3893 & 27.0481 & 0.1153 & 3442 & 22.8496 & 22.2698 & 22.9225 & 22.2908 \\
    0.0652 & 1197 & 29.1869 & 26.9548 & 29.2622 & 26.9718 & 0.1243 & 3513 & 22.6326 & 22.0612 & 22.7055 & 22.0822 \\
    0.0656 & 1255 & 28.9354 & 26.8050 & 29.0104 & 26.8220 & 0.1338 & 3618 & 22.3916 & 21.8293 & 22.4644 & 21.8505 \\
    0.0658 & 1281 & 28.8239 & 26.7377 & 28.8988 & 26.7548 & 0.1427 & 3655 & 22.2265 & 21.6670 & 22.2993 & 21.6882 \\
    0.0661 & 1345 & 28.4528 & 26.6167 & 28.5319 & 26.6339 & 0.1542 & 3688 & 22.0459 & 21.4885 & 22.1186 & 21.5097 \\
    0.0662 & 1362 & 28.3422 & 26.5901 & 28.4230 & 26.6073 & 0.1693 & 3721 & 21.8433 & 21.2856 & 21.9160 & 21.3069 \\
    0.0664 & 1418 & 28.0347 & 26.4920 & 28.1192 & 26.5094 & 0.1869 & 3755 & 21.6378 & 21.0787 & 21.7104 & 21.0999 \\
    0.0666 & 1470 & 27.8727 & 26.3837 & 27.9570 & 26.4012 & 0.2072 & 3789 & 21.4319 & 20.8716 & 21.5045 & 20.8929 \\
    0.0669 & 1533 & 27.7282 & 26.2735 & 27.8119 & 26.2911 & 0.2304 & 3822 & 21.2287 & 20.6682 & 21.3013 & 20.6895 \\
    0.0669 & 1561 & 27.6830 & 26.2303 & 27.7662 & 26.2480 & 0.2573 & 3860 & 21.0215 & 20.4610 & 21.0940 & 20.4824 \\
    0.0671 & 1646 & 27.4360 & 26.0711 & 27.5190 & 26.0890 & 0.2880 & 3916 & 20.7549 & 20.1950 & 20.8274 & 20.2165 \\
    0.0671 & 1648 & 27.4271 & 26.0665 & 27.5102 & 26.0844 & 0.3154 & 3993 & 20.4918 & 19.9359 & 20.5643 & 19.9573 \\
    0.0673 & 1729 & 27.1148 & 25.8841 & 27.1981 & 25.9021 & 0.3397 & 4064 & 20.2342 & 19.7167 & 20.3067 & 19.7381 \\
    0.0673 & 1736 & 27.0898 & 25.8657 & 27.1730 & 25.8837 & 0.3627 & 4167 & 19.9717 & 19.4819 & 20.0441 & 19.5034 \\
    0.0675 & 1800 & 26.8668 & 25.7019 & 26.9490 & 25.7200 & 0.3781 & 4266 & 19.7909 & 19.3138 & 19.8634 & 19.3354 \\
    0.0677 & 1864 & 26.6838 & 25.5681 & 26.7654 & 25.5863 & 0.3831 & 4300 & 19.7320 & 19.2599 & 19.8045 & 19.2815 \\
    0.0681 & 1931 & 26.5027 & 25.4430 & 26.5839 & 25.4613 & 0.3980 & 4417 & 19.5631 & 19.1111 & 19.6357 & 19.1328 \\
    0.0685 & 2010 & 26.3045 & 25.3151 & 26.3856 & 25.3335 & 0.4107 & 4527 & 19.4259 & 18.9931 & 19.4988 & 19.0148 \\
    0.0691 & 2117 & 26.0177 & 25.1107 & 26.0980 & 25.1293 & 0.4222 & 4634 & 19.3078 & 18.8934 & 19.3810 & 18.9151 \\
    0.0697 & 2206 & 25.7925 & 24.9398 & 25.8719 & 24.9586 & 0.4330 & 4741 & 19.2020 & 18.8059 & 19.2756 & 18.8276 \\
    0.0700 & 2243 & 25.7030 & 24.8708 & 25.7820 & 24.8896 & 0.4431 & 4845 & 19.1057 & 18.7268 & 19.1797 & 18.7486 \\
    0.0704 & 2287 & 25.5962 & 24.7882 & 25.6748 & 24.8072 & 0.4528 & 4946 & 19.0131 & 18.6508 & 19.0875 & 18.6726 \\
    0.0711 & 2377 & 25.3842 & 24.6165 & 25.4619 & 24.6357 & 0.4624 & 5048 & 18.9217 & 18.5753 & 18.9965 & 18.5970 \\
    0.0720 & 2463 & 25.1840 & 24.4591 & 25.2611 & 24.4785 & 0.4718 & 5148 & 18.8303 & 18.4990 & 18.9054 & 18.5208 \\
    0.0730 & 2549 & 24.9889 & 24.2942 & 25.0652 & 24.3137 & 0.4812 & 5247 & 18.7376 & 18.4209 & 18.8130 & 18.4427 \\
    0.0742 & 2635 & 24.7996 & 24.1246 & 24.8753 & 24.1443 & 0.4907 & 5347 & 18.6423 & 18.3397 & 18.7181 & 18.3614 \\
    0.0756 & 2719 & 24.6150 & 23.9581 & 24.6901 & 23.9779 & 0.5002 & 5445 & 18.5431 & 18.2539 & 18.6191 & 18.2757 \\
    0.0764 & 2761 & 24.5244 & 23.8764 & 24.5993 & 23.8964 & 0.5099 & 5543 & 18.4382 & 18.1620 & 18.5145 & 18.1838 \\
    0.0766 & 2771 & 24.5026 & 23.8567 & 24.5774 & 23.8767 & 0.5198 & 5639 & 18.3253 & 18.0617 & 18.4019 & 18.0835 \\
    0.0783 & 2847 & 24.3371 & 23.7016 & 24.4115 & 23.7217 & 0.5301 & 5736 & 18.2007 & 17.9492 & 18.2775 & 17.9710 \\
    0.0804 & 2931 & 24.1609 & 23.5318 & 24.2349 & 23.5521 & 0.5407 & 5830 & 18.0590 & 17.8191 & 18.1361 & 17.8409 \\
    0.0830 & 3014 & 23.9881 & 23.3654 & 24.0617 & 23.3859 & 0.5520 & 5924 & 17.8916 & 17.6629 & 17.9690 & 17.6847 \\
        \cmidrule(lr){1-6} \cmidrule(lr){7-12}
    \end{tabular}
    \label{tab:HMHA40}
\end{table*}

\end{appendix}

\end{document}